\documentclass[aps,pra,twocolumn,10pt,superscriptaddress]{revtex4-2}
\usepackage[utf8]{inputenc}
\usepackage{graphicx}
\usepackage{hyperref}
\hypersetup{colorlinks = true, linkcolor = [rgb]{0.19411,0.51882,0.667058}, urlcolor = [rgb]{0.125490,0.29542,0.1647058}, citecolor = [rgb]{0.75882,0.37411,0.14117}}
\usepackage{amsmath}
\usepackage{blkarray}
\usepackage{amsfonts}
\usepackage{amssymb}
\usepackage{siunitx}
\usepackage{braket}
\usepackage{import}
\usepackage{dsfont}
\usepackage{color,soul}
\usepackage{physics}
\usepackage[caption=false]{subfig}

\renewcommand{\paragraph}[1]{\addcontentsline{toc}{section}{#1}\emph{#1.}---}
\newcommand*{\avg}{\mathrm{avg}}
\newcommand*{\rel}{\mathrm{rel}}
\newcommand*{\quant}{\mathrm{QT}}
\newcommand*{\taylor}{\mathrm{approx}} 
\newcommand*{\fit}{\mathrm{Fit}}
\newcommand*{\newton}{\mathrm{N}}

\begin{document}

\title{Newton's laws of motion generating gravity-mediated entanglement}

\author{Marta M. Marchese}
\affiliation{Naturwissenschaftlich-Technische Fakult\"{a}t, Universit\"{a}t Siegen, Walter-Flex-Stra\ss e 3, 57068 Siegen, Germany}

\author{Martin Pl\'{a}vala}
\affiliation{Naturwissenschaftlich-Technische Fakult\"{a}t, Universit\"{a}t Siegen, Walter-Flex-Stra\ss e 3, 57068 Siegen, Germany}
\affiliation{Institut f\"{u}r Theoretische Physik, Leibniz Universit\"{a}t Hannover, Hannover, Germany}

\author{Matthias Kleinmann}
\affiliation{Naturwissenschaftlich-Technische Fakult\"{a}t, Universit\"{a}t Siegen, Walter-Flex-Stra\ss e 3, 57068 Siegen, Germany}

\author{Stefan Nimmrichter}
\affiliation{Naturwissenschaftlich-Technische Fakult\"{a}t, Universit\"{a}t Siegen, Walter-Flex-Stra\ss e 3, 57068 Siegen, Germany}

\begin{abstract}
The interface between quantum theory and gravity represents still uncharted territory. Recently, a tabletop experiment has been proposed for witnessing quantum features of gravity: Two masses in an initial superposition of spatially localized states interact only through gravity and it is measured whether the final state is entangled. Here we show that one can generate the same amount of entanglement in this setup by using the time evolution given by Newton's laws of motion. We argue that theories of quantum gravity that can be approximated by the Newtonian potential and time evolution given by Newton's laws of motion will generate gravity-mediated entanglement.
\end{abstract}

\maketitle

\paragraph{Introduction}
Whether gravity is fundamentally quantum or not is still one of the most debated problems since the lack of experimental evidence does not allow for a decisive answer in either direction. Gravity-mediated entanglement has recently been suggested as a testable quantum feature to address this question~\cite{bose2017spin,marletto2017gravitationally,al2018optomechanical,krisnanda2020observable,weiss2021large}. In these schemes, two masses are prepared in a superposition of spatial states on the two arms of an interferometer, then they are let to interact through just the gravitational field. If gravity-mediated entanglement between the two masses is detected at the end of the evolution, then an argument that local operations and classical communications (LOCC) cannot generate entanglement~\cite{horodecki2009quantum,guhne2009entanglement} is used to conclude that the interaction cannot be described by classical physics: either gravity is non-local, or it is nonclassical~\cite{galley2022no, marshman2020locality}. In this sense the proposed experiments aim to answer the standing question whether gravity needs to be quantized or not~\cite{tilloy2019does,oppenheim2023time}. Performing the proposed experiment is still outside the realm of possibility~\cite{aspelmeyer2022zeh}, but many variants of the scheme \cite{nguyen2020entanglement,lami2023testing,howl2023gravitationally,hanif2023testing, pesci2023testing,pedernales2022enhancing,das2024mass,christodoulou2023gravity,overstreet2023inference,bengyat2023gravity} as well as additional experiments \cite{hilweg2017gravitationally,qvarfort2022constraining,panda2023measuring} were recently proposed and additional theoretical analysis was carried out to further support the previous results~\cite{danielson2022gravitationally,christodoulou2023locally,feng2023conservation,bose2022mechanism}.

\begin{figure}
 \centering
 \includegraphics[width=\linewidth]{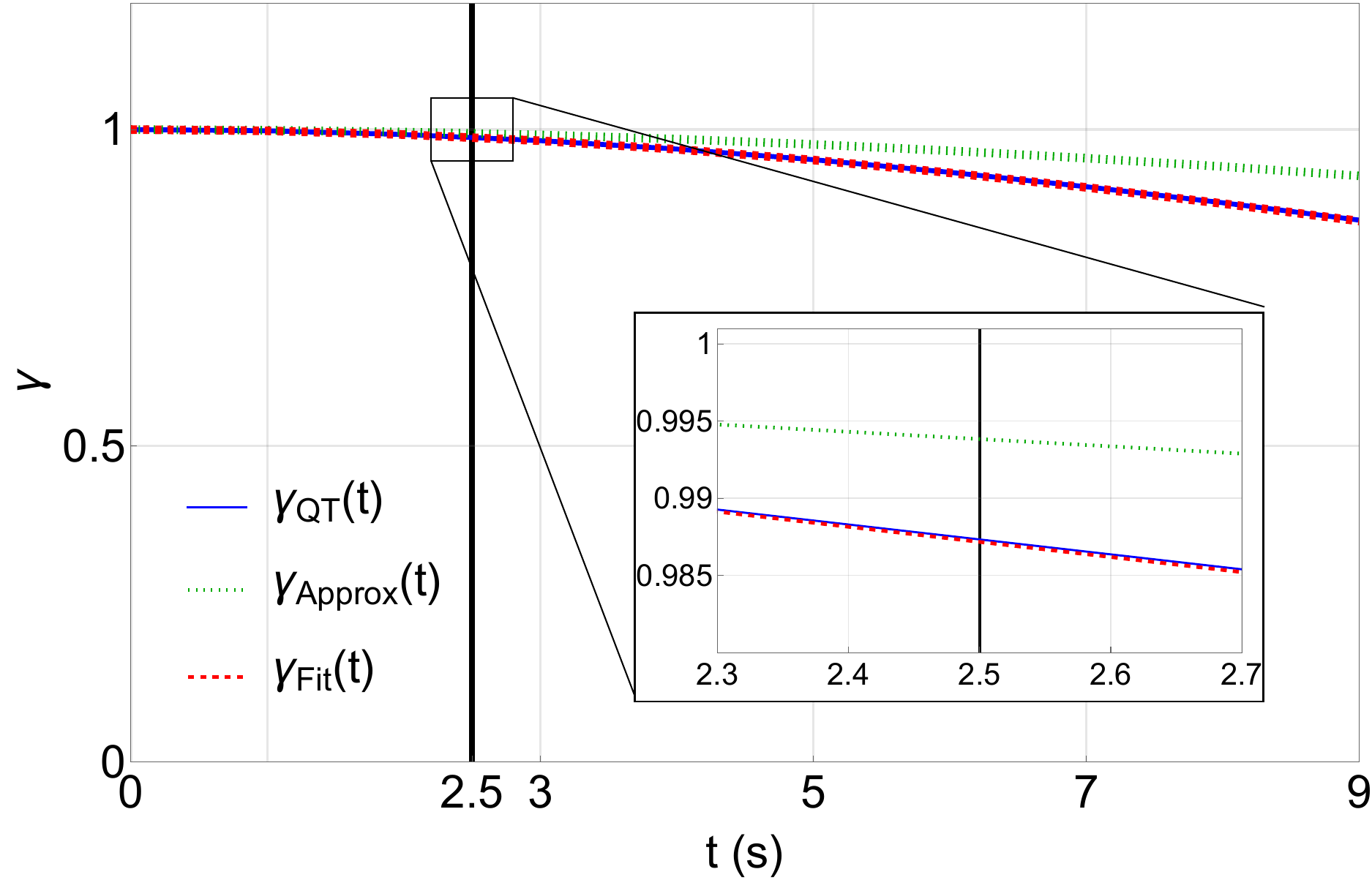}
 \caption{ Purity marginal as a function of time. $\gamma_\quant$ (blue solid line) is the purity of the marginal of the state evolved according to the Schr\"{o}dinger equation; $\gamma_\taylor$ (green dotted line) is the purity of the marginal obtained from the time evolution in the phase space of the Wigner function by using the second-order Taylor expansion of the potential; $\gamma_\fit$ (red dashed line) is the purity for the marginal obtained from the time evolution with the second-order polynomial that coincides with the Newtonian potential for $x_2 - x_1 = d \pm \Delta x$ and $x_2 - x_1 = d$. Here $x_1$ and $x_2$ are the position coordinates of the two particles, $d$ is the distance of centers of mass of the superpositions and $\Delta x$ is the arm separation. We are using parameters as proposed in Ref.~\cite{bose2017spin}, i.e., $m = 10^{-14} \text{kg}$, $\Delta x = 250 \mu \text{m}$, $d = 450 \mu \text{m}$, and $\sigma=10 \mu \text{m}$. The solid black line indicates the proposed time $t=2.5\text{s}$ \cite{bose2017spin} over which the phases $\varphi_1=-0.13$ and $\varphi_2=0.44$ are accumulated.}
 \label{fig:purity}
\end{figure}

Witnessing gravity-mediated entanglement will rule out several alternative theories such as semi-classical models, that treat gravity as a classical field acting as a weak continuous measurement on a quantum system~\cite{kafri2014classical,Khosla2018,oppenheim2023postquantum,tilloy2019does,Carney2023}, and theories predicting the breaking down of quantum mechanics at scales where quantum gravity effects should be relevant~\cite{diosi1989models,tilloy2016sourcing}. Along the same line, some measurement-feedback descriptions have been ruled out because they are incompatible with experimental data from atom interferometry experiments~\cite{altamirano2018gravity}. However, there are still classical models that have not been excluded yet~\cite{tilloy2019does,oppenheim2023postquantum}.

It has also been argued that witnessing gravity-induced entanglement does not undoubtedly certify the nonclassical nature of gravity~\cite{hall2018two,anastopoulos2021gravitational,fragkos2022inference,doner2022gravitational,spaventa2023tests,foo2023relativity}. For the electromagnetic interaction in the Lorenz gauge the entanglement is mediated by non-physical scalar and longitudinal photons \cite{franson2011entanglement,fragkos2022inference} and for the gravitational interaction in the Lorenz gauge the entanglement is mediated by the virtual gravitons \cite{marshman2020locality,bose2022mechanism,fragkos2022inference}. These results cast doubt on whether the LOCC argument can be used on non-physical objects as, apart from locality, a key assumption in the LOCC argument is that the interaction is mediated by a physical classical system. For example, one can assign negative probabilities to the creation or detection of virtual exchange particles \cite{Feynman1987-FEYNP} and it was recently demonstrated that one can violate Bell inequalities \cite{brunner2014bell} with local indirect access to negative probabilities \cite{onggadinata2023simulations}; it is well-known that Bell inequalities cannot be violated in the LOCC regime and thus the LOCC argument cannot be applied. Moreover, using essentially the contrapositive of the LOCC argument, the observation of gravity-mediated entanglement should prove that the gravitational field was at some point entangled with the particles, but this is impossible since the state of the particles is always pure even if one describes the experiment using a first-principles approach and path integrals \cite{christodoulou2023locally}. This is a clear contradiction since monogamy of entanglement states that a pure state cannot be entangled with any other state, even if one would consider the gravitational field to be described by an operational theory potentially different from quantum theory \cite{plavala2023general}. Another argument that gravity-mediated entanglement proves quantization of gravity is that it requires the virtual gravitons to be mathematically described by operators \cite{marshman2020locality,bose2022mechanism} but this is again a problematic argument: systems described by operators can be classical in the sense that non-contextual model exists \cite{kochen1990problem,spekkens2005contextuality,abramsky2011sheaf,raussendorf2013contextuality,dzhafarov2017contextuality,mansfield2018quantum,budroni2022kochen} and one can also describe the classical mechanics in Hilbert space by using the Koopman-von Neumann construction \cite{koopman1931hamiltonian,neumann1932operatorenmethode,neumann1932zusatze,piasecki2021introduction}. Nevertheless, as also pointed out in Ref.~\cite{fragkos2022inference}, one can always describe the experiment in the Coulomb gauge using the non-local potentials and one can look for non-classical effects in this description. This circumvents the need for a consistent, and so far lacking, theory of ``gravitons''.
What is then a criterion for nonclassicality in the Coulomb gauge? Here, quantum theory enters the description of the experiment in two ways: first, the initial state of the particles is a quantum superposition, and second, the time evolution under the gravitational potential is given by the Schr\"{o}dinger equation. Let us assume that the latter holds for non-gravitational interactions and that the preparation of the initial superposition state does not involve gravity. Then the minimal criterion for nonclassicality of the gravitational interaction is to observe dynamical effects arising from its contribution to the Schr\"{o}dinger equation that cannot be explained via a suitable classical approximation of the time evolution.

To be more specific, the Schr\"{o}dinger equation can be equivalently formulated in phase space via the Moyal bracket \cite{moyal1949quantum,groenewold1946principles} which can be expanded in powers of $\hbar^2$. The first term of the Moyal bracket, of the order $\hbar^0$, is identical to the Poisson bracket that describes the classical time evolution in phase space based on Newton's laws of motion. We thus propose the following non-classicality criterion:
in order to probe non-classical dynamical effects (of the gravitational interaction) in an experiment, quantum theory must predict an observable contribution from (gravitational) terms in the Moyal bracket with non-zero order of $\hbar^2$. It is straightforward to check whether this is the case: one compares the prediction based on the Moyal bracket including only the (gravitational) $\hbar^0$-terms to the one based on the full Schr\"odinger equation. If the predictions coincide, the experiment cannot rule out classical dynamics where the interaction is mediated solely by the gravitational potential, though classical models based on exchange particles can still be ruled out. This approach has the caveat that one has to take into account that letting a quantum state evolve according to a truncated Moyal bracket may result in unphysical states, for example, predicting negative probabilities.

Here we show that it is possible to reproduce the gravity-mediated entanglement between the two simultaneously interfering masses by using a 
time evolution based on Newton's laws of motion and a suitable approximation of the gravitational potential thus showing that non-classical dynamical effects are not present. The amount of entanglement, quantified using the purity of the marginals, matches the quantum prediction based on the Schr\"{o}dinger equation, see Fig.~\ref{fig:purity}. This has two implications: Firstly, any theory of gravity, for which in the regime of tabletop experiments Newton's laws of motion are the leading term of the time evolution in phase space, is capable of generating entanglement from spatial superpositions. Secondly, the aforementioned experimental test schemes will not provide a definitive answer about the quantum nature of gravity, since they do not rule out a simple classical description.

We would like to clarify how we intend the meaning of classicality in this framework. We do not suggest a classical description of gravity, in the sense that we do not consider gravity mediated by a classical field. Rather, we restrict the time-evolution to be classical in the sense that we take the limit ``$\hbar\to 0$'' for the dynamical equation. We observe that the prediction under this notion of classicality cannot be distinguished from the dynamics governed by the Schrödinger equation.

In the same way as Ehrenfest's theorem provides a classical description of a quantum system in terms of operator expectation values, here, we show that the current proposed experimental setups can be described in terms of Liouville equations and that the entanglement predictions are equivalent. Therefore, our results point out a crucial distinction between the non-classicality (intended as entanglement production) arising from the dynamics of a system evolving under gravitational potential and the quantization of the mediators of gravity. These are two independent aspects. The generation of entanglement is only one aspect for a theory to be non-classical, it is not necessary in the sense that there are non-classical operations that do not generate entanglement (e.g., swap operation), and it is not sufficient in the sense that there are systems that develop entanglement under the evolution of a simply non-local potential. There are more requirements, beyond the simple generation of entanglement, that a quantum evolution should obey to infer something about the underlying nature of the process. We propose to investigate the discrepancy that arises in the dynamics between Poisson and full Moyal brackets, as a different requirement for a dynamics to be non-classical.

In the following, we treat the two gravitationally interacting masses in the nonrelativistic Newtonian limit. To calculate the time evolution of a quantum state according to Newton's laws of motion, we employ the Wigner function representation~\cite{case2008wigner} and the Liouville equation. We also use a second-order polynomial approximation of the interaction potential, which yields an analytically tractable and physical solution of the masses' state that exhibits entanglement without making any assumption on the quantum nature of gravity. While this approximation may seem crude, it predicts the result of the experiment extremely well on the proposed time scale of the experiment. Moreover, it always leads to a well-defined quantum state since for Hamiltonians that are polynomials up to second order the classical and quantum time evolution coincide. Other approximations of the potential will generally
result in unphysical Wigner functions over time.
For example, a stepwise linear approximation, subjecting different branches of the superposition to different gravitational forces, leads to small but noticeable negative probabilities at $t\simeq 2.5\;\text{s}$  for the settings of Fig.~\ref{fig:purity}; this could be remedied by an adequate amount of diffusion noise, which would arise in a consistent hybrid model of quantum particles interacting with a classical gravitational field \cite{tilloy2019does, oppenheim2023postquantum}. However, this noise also prevents the detection of entanglement using the outlined methods, see Appendix~\ref{appendix:stepwisepotential}.

We proceed by briefly reviewing the interferometric setup and the wavefunction description of the proposed quantum gravity test schemes and introduce the purity-based entanglement measure we employ. We then present our classical phase-space model for gravity-mediated entanglement underlying the results plotted in Fig.~\ref{fig:purity} and conclude with an outlook on the future of such test experiments.

\paragraph{Experimental scenario}
Consider a Mach-Zehnder interferometer setup with two neutral identical masses $m$ placed at a distance $d$. Each of them is split into a superposition of spatial states $\ket{L}$ and $\ket{R}$ representing the two arms of the interferometer with a separation $\Delta x < d$. Each single-arm state is given by a localized Gaussian wave packet of width $\sigma \ll \Delta x$. Hence, the two masses are initially in a normalized product state $\psi(x_1,x_2)=\psi_1(x_1) \psi_2(x_2)$ of two spatially superimposed Gaussian wave packets,
\begin{equation}
\psi_i(q_i) = \dfrac{e^{-(q_i - \Delta x/2)^2/4\sigma^2} + e^{-(q_i + \Delta x/2)^2/4\sigma^2}}{(2\pi\sigma^2 N)^{1/4}}, 
\label{eq:gaussiansuperposition}
\end{equation}
where $q_1=x_1 - d/2$ and $q_2=x_2 + d/2$ are the shifted one-dimensional coordinates of the two masses and $N = 4(1+e^{-\Delta x^2/8\sigma^2})^2$ is the normalization constant. We shall omit the other two spatial directions throughout, assuming that both masses are sufficiently localized in $y,z$ at all times. We assume that the distance between the two masses is always sufficiently large so that we can neglect the Casimir-Polder force, and the only force acting between the two masses is due to their mutual gravitational attraction. The Hamiltonian for the two masses is then $H(x_1, x_2, p_1, p_2) = (p_1^2 + p_2^2)/2m - \kappa / \abs{x_1 - x_2}$, with $\kappa=G m^2$. We follow the calculations presented in \cite{bose2017spin}: The two Gaussian wave packets now correspond to the left and right arms of an interferometer represented by the states $\ket{L}$ and $\ket{R}$. Then, using the full Newtonian potential and neglecting the kinetic part of the Hamiltonian, the time evolution due purely to gravitational interaction leads to the state 
\begin{equation}
\begin{split}
\ket{\psi(t)} = \dfrac{e^{i \phi}}{\sqrt{N(t)}} &(\ket{LL} + e^{i\Delta \phi_{LR} t} \ket{LR} \\
&+ e^{i\Delta \phi_{RL} t} \ket{RL} + \ket{RR} ), \label{eq:quantumstate}
\end{split}
\end{equation}
where $N(t)$ is a normalization factor that depends on the phase shifts if the states are not perfectly orthogonal.
For $\Delta x \gg \sigma$ we get $N(t) = 4$ which we will use from now on. The phases $\Delta \phi_{LR} = \phi_{LR} - \phi$, $\Delta \phi_{RL} = \phi_{RL} - \phi$ read $\phi_{LR} = \frac{G m^2}{\hbar(d + \Delta x)}$, $\phi_{RL} = \frac{G m^2}{\hbar(d - \Delta x)}$, $\phi = \frac{G m^2}{\hbar d}$. The final state is still pure, since the evolution is unitary. In order to check if it is still separable or not we look at the purity of marginals. In particular, the marginal state of particle $2$ is given by tracing out particle 1, $\rho_2(t) = \Tr_1[\rho(t)]$, where $\rho(t) = \dyad{\psi(t)}$ is the density matrix of the evolved state from Eq.~\eqref{eq:quantumstate}. The reduces states density matrix results in
\begin{equation}
\rho_2(t) = \dfrac{1}{4}
\begin{pmatrix}
2 & e^{-i\Delta \phi_{LR} t} - e^{i \Delta \phi_{RL} t}\\
e^{i \Delta \phi_{LR} t} - e^{- i \Delta \phi_{RL} t} & 2
\end{pmatrix}
\end{equation}
and it is straightforward to obtain the purity
\begin{equation}
\gamma_\quant(t) = \Tr[\rho_2^2(t)]=\dfrac{3 + \cos( (\Delta \phi_{LR} + \Delta \phi_{RL}) t )}{4}.
\end{equation}
If a quantum state represented by a density matrix $\rho$ is pure, a necessary and sufficient condition for separability is that all the marginals must also be pure. Therefore, since the purity of the marginal is found smaller than $1$, then this implies that $\ket{\psi(t)}$ must be entangled.

There are two ways in which non-classical effects enter the previous calculations: the initial states are assumed to be superpositions of paths and we use the Sch\"{o}dinger equation to evolve the state in time. If we changed the initial states to statistical mixtures instead of superpositions then entanglement would not be generated, hence the non-classical nature of the initial state is crucial for obtaining gravity-mediated entanglement. The question we aim to answer is: what would happen if we changed the quantum time evolution given by Sch\"{o}dinger equation to a time evolution given by Newton's laws of motion? This is a relevant question since if one can generate entanglement using a classical time evolution, then the only necessary non-classical resource are the initial superpositions.

\paragraph{Newton's laws of motion evolution}
In phase space, Newton's laws of motion correspond to the Liouville equation $\dot{P} = \{H, P\}$, where $\{\cdot, \cdot\}$ is the Poisson bracket, $\{f,g\} = \frac{\partial f}{\partial q} \frac{\partial g}{\partial p} - \frac{\partial f}{\partial p} \frac{\partial g}{\partial q}$, $H$ is the Hamiltonian, and $P$ is the probability distribution of the system in phase space. In order to apply Newton's laws of motion to a quantum system, we simply replace the probability distribution $P$ by its closest quantum mechanical analog, the Wigner function $W$ \cite{schleich2011quantum, plavala2022operational}. The Wigner function for a single particle in a pure state is
\begin{equation}
W(q,p) = \dfrac{1}{2\pi\hbar} \int e^{-i ps/\hbar} \psi^{*} \left(q-\frac{s}{2}\right)\psi \left(q+\frac{s}{2} \right) \dd s. 
\end{equation}
and we get the equation
\begin{equation}
\dot{W} = \{H, W\} \label{eq:wigDot}
\end{equation}
where $H$ is still the same Hamiltonian function in phase space. As it is the case for the Liouville equation, also Eq.~\eqref{eq:wigDot} is solved by $W(x_1,x_2,p_1,p_2,t) = W^0 (X_1(t), X_2(t), P_1(t), P_2(t))$, where $X_i(t) = X_i(x_1, x_2, p_1, p_2, t)$ and $P_i(t) = X_i(x_1, x_2, p_1, p_2, t)$ are the coordinates of the point in phase space at which pair of classical particles would have to start at $t = 0$ to reach the point $(x_1,x_2,p_1,p_2)$ at time $t$. It now only remains to solve the respective Hamilton equations of motion.

At the initial time $t=0$, the total Wigner function for the two-particle state will be given by the product 
\begin{equation}
W^0(x_1,x_2,p_1,p_2) = W_1^0(x_1,p_1) W^0_2(x_2,p_2), \label{eq:wig0}
\end{equation}
where we used Eq.~\eqref{eq:gaussiansuperposition} to obtain the explicit expression for each of the two masses,
\begin{equation}
\begin{split}
 W_i^0(q_i,p_i) = \; &\dfrac{e^{\frac{-2\sigma^2 p_i^2}{\hbar^2}}}{\pi \hbar \sqrt{N}} \Bigg[ e^{-\frac{(q_i+\frac{\Delta x}{2})^2}{2\sigma^2}}+ e^{-\frac{(q_i-\frac{\Delta x}{2})^2}{2\sigma^2}} \\
 &+ 2 e^{-\frac{q_i}{2\sigma^2}} \cos(\frac{\Delta x \; p_i}{\hbar}) \Bigg].
\end{split} \label{eq:singleparticleW}
\end{equation}
see Appendix~\ref{appendix:initial} for detailed calculations.

In order to make the problem tractable, we will approximate the Newtonian gravitational potential by a second-order polynomial. This ensures that the time-evolution yields a pure quantum state at all times \cite{case2008wigner}. To do so, we first apply a symplectic transformation that preserves both the Poisson bracket and the commutator. The new coordinates are
\begin{equation}
\begin{split}
x_\avg = \dfrac{1}{\sqrt{2}}(x_1 + x_2), \qquad
&p_\avg = \dfrac{1}{\sqrt{2}}(p_1 + p_2), \\
x_\rel = \dfrac{1}{\sqrt{2}}(x_2 - x_1 - d), \qquad
&p_\rel = \dfrac{1}{\sqrt{2}}(p_2 - p_1),
\end{split} \label{eq:newCoordinates}
\end{equation}
and the Hamiltonian is
\begin{equation}\label{eq:Hamiltonian}
H(x_\avg, x_\rel, p_\avg, p_\rel) = \dfrac{p_\avg^2 + p_\rel^2}{2m} - \dfrac{\kappa}{d + \sqrt{2} x_\rel}.
\end{equation}
Here, we assumed that during the whole experiment, the Wigner function approximately vanishes at $x_1 \geq 0$ or at $x_2 \leq 0$. Therefore, $d + \sqrt{2} x_\rel \geq 0$ and so we dropped the absolute value.

The Taylor expansion of the potential up to second order in $x_\rel$ is
\begin{equation}
V_\taylor(x_\rel) = - \dfrac{\kappa}{d} + \dfrac{\sqrt{2} \kappa}{d^2} x_\rel - \dfrac{2 \kappa}{d^3} x_\rel^2, \label{eq:VTaylor}
\end{equation}
Alternatively, we will consider a quadratic fit of the Newtonian potential such that the fitting polynomial coincides with the potential at the most relevant points $x_\rel = \pm \Delta x/\sqrt{2}$ and $x_\rel = 0$,
\begin{equation}
V_\fit(x_\rel) = - \dfrac{\kappa}{d} + \dfrac{\sqrt{2} \kappa}{d^2 - \Delta x^2} x_\rel - \dfrac{2 \kappa}{d ( d^2 - \Delta x^2 )} x_\rel^2. \label{eq:Vfit}
\end{equation}
Both of the quadratic potentials $V_\taylor$ and $V_\fit$ give reasonable approximations to the Newtonian potential and the respective force, see Appendix~\ref{appendix:approx}. In both cases, the respective Hamilton equations of motion can be easily solved, see Appendix~\ref{appendix:timeEvo}. The two approximations will in general lead to different time evolutions, but note that for $d \gg \Delta x$ we have $V_\taylor(x_\rel) \approx V_\fit(x_\rel)$. See Appendix \ref{appendix:benchmarktraj} for a benchmark comparison against the exact 
time evolution. We will use $W^\taylor$ and $W^\fit$ to denote the Wigner functions evolved according to Hamiltonians with the approximate potentials $V_\taylor$ and $V_\fit$, respectively.

We will again use purity of the marginals to quantify entanglement of the states; we use this method since it is easy to compute from the Wigner function \cite{almeida2009entanglement}. We want to compute the quantities
\begin{align}
\gamma_\taylor(t) &= 2\pi\hbar \int (W^\taylor_2(x_2, p_2, t))^2 \dd x_2 \dd p_2 \\
\gamma_\fit(t) &= 2\pi\hbar \int (W^\fit_2(x_2, p_2, t))^2 \dd x_2 \dd p_2
\end{align}
where
\begin{align}
W^\taylor_2(x_2, p_2, t) &= \int W^\taylor(x_1, x_2, p_1, p_2, t) \dd x_1 \dd p_1 \\
W^\fit_2(x_2, p_2, t) &= \int W^\fit(x_1, x_2, p_1, p_2, t) \dd x_1 \dd p_1
\end{align}
are the respective marginals. Since we have already solved the time evolution, we can compute the purities numerically.

In Fig.~\ref{fig:purity} we show the purity of the reduced state as a function of time for the three different time evolutions: quantum dynamics given by Schr\"{o}dinger equation and classical dynamics given by Newton's laws of motion and the Taylor approximated potential or the second-order polynomial fit of the potential. The state is pure at the initial time $t = 0$, since the total state was just the product of the two particle states. 
We focus on a time span of less than 10 seconds in which the deflection and dispersion of the wavepackets can be safely neglected and thus the quantum model in Eq.~\eqref{eq:quantumstate} remains valid. Throughout, the second-order fit predicts almost exactly the same entanglement as the quantum model, whereas the Taylor approximation deviates noticeably after a few seconds.

\paragraph{Conclusions}
In this paper we have shown that gravity-mediated entanglement can be generated by Newton's laws of motion, here represented via the Liouville equation using a suitable approximation of the Newtonian potential. Hence, observing gravity-mediated entanglement does not certify the non-classicality of the gravitational interaction, in the sense that initial superpositions and second-order interaction potential are already sufficient, when the time-evolution is approximated by the dynamical equation in the classical limit $\hbar\to 0$.

Detection of entanglement in these experiments can be used to exclude a wide range of classical theories of gravity, namely the ones that fall in the LOCC category, where the interaction is assumed to be mediated by a local classical mediator. However, the set of potential theories that are non-LOCC is still broad, and we showed that it will not be possible to discriminate these theories by simply looking at the entanglement production in upcoming experiments. Instead, we suggest to study effects arising from the quantum terms proportional to $\hbar^2$ in the Moyal brackets and design new experiments that are sensitive to such deviations from the Liouville equation for gravity.
  
For example, one could consider different arm separations $\Delta x$ for the two particles, which would invalidate a second-order approximation of the gravitational potential $V_\fit$ given by \eqref{eq:Vfit}, and one could let the masses interact for sufficiently long times, which would invalidate the approximation $V_\taylor$ given by \eqref{eq:VTaylor}
by leading to different predictions for the entanglement rate in the setup of Ref.~\cite{bose2017spin}.
An even more conclusive, but much more challenging, version of the experiment could use the gravitational interaction to create the initial superpositions.

In order to tackle the problem more systematically,
an approach based on operational theories~\cite{plavala2022operational,plavala2023generalized} similar to the approach to violations of Bell inequalities can be adopted:  This approach was recently used to prove that quantum theory based on real numbers can be experimentally ruled out~\cite{renou2021quantum}, to analyze quantumness of uncertainty relations \cite{catani2022nonclassical} and interference phenomena \cite{blasiak2015local,catani2023aspects,catani2023interference}. A similar approach to quantum gravity may be able to rule out large classes of classical descriptions and settle the question of whether gravity is non-classical. We have already made a first step in this direction in this work by showing that, in the current experimental setups, an approximation to the second-order of the gravitational potential is enough to obtain the same results as the quantum description, without making assumptions on the nature of gravity, which could still in principle be classical.

A potential criticism of our approach is that we are using the Newtonian potential to model the gravitational interaction which is by definition nonlocal. This is the standard approach in the quantum description of the proposed experiments \cite{bose2017spin,marletto2017gravitationally,al2018optomechanical,krisnanda2020observable,weiss2021large}, and we are here concerned on whether this quantum description involves genuine quantum dynamics driven solely by the gravitational potential. In this regime, particle exchange does not play a role, so we do not base our argument on it.

\paragraph{Acknowledgments}
We acknowledge support from the Deutsche Forschungsgemeinschaft (DFG, German Research Foundation, project numbers 447948357 and 440958198), the Sino-German Center for Research Promotion (Project M-0294), the German Ministry of Education and Research (Project QuKuK, BMBF Grant No. 16KIS1618K), the DAAD.
MP acknowledges support from the Alexander von Humboldt Foundation. MMM acknowledges support from the Walter Benjamin Programme (project number 510053905).

\onecolumngrid
\appendix

\section{Wigner function for the initial state} \label{appendix:initial}
The Wigner function for a single-particle quantum state described by a wavefunction $\psi(q)$ is
\begin{equation}
W(q,p) = \dfrac{1}{2 \pi \hbar} \int e^{-i \frac{ps}{\hbar}} \psi^{*}(q - \frac{s}{2})\psi(q + \frac{s}{2}) \dd s.
\end{equation}
Then if we have Gaussian wavepackets of the form
\begin{equation}
\psi(q) = \left(\dfrac{1}{2\pi\sigma^2 N}\right)^{1/4}\left[e^{-\frac{1}{4\sigma^2}\left(q+\frac{\Delta x}{2}\right)^2}+e^{-\frac{1}{4\sigma^2}\left(q-\frac{\Delta x}{2}\right)^2}\right], \qquad \text{with }\,\, N=4\left[1+e^{-\frac{(\Delta x)^2}{8\sigma^2}}\right]^2,
\end{equation}
we obtain
\begin{equation}
\begin{split}
W(q,p)=\dfrac{1}{2\pi\hbar}\left(\dfrac{1}{2\pi \sigma^2N}\right)^{1/2} \int e^{-i \frac{ps}{\hbar}} &\Bigg[e^{-\frac{1}{4\sigma^2}\left(q+\frac{\Delta x -s}{2}\right)^2-\frac{1}{4\sigma^2}\left(q+\frac{\Delta x +s}{2}\right)^2}\\
&+e^{-\frac{1}{4\sigma^2}\left(q+\frac{\Delta x-s}{2}\right)^2-\frac{1}{4\sigma^2}\left(q-\frac{\Delta x-s}{2}\right)^2}\\
&+e^{-\frac{1}{4\sigma^2}\left(q-\frac{\Delta x -s}{2}\right)^2-\frac{1}{4\sigma^2}\left(q+\frac{\Delta x +s}{2}\right)^2}\\
&+e^{-\frac{1}{4\sigma^2}\left(q-\frac{\Delta x+s}{2}\right)^2-\frac{1}{4\sigma^2}\left(q-\frac{\Delta x-s}{2}\right)^2}\Bigg] \dd s.
\end{split}
\end{equation}
The integral will yield a real function in the end, such that the result for one particle is
\begin{equation}
 W(q,p)=\dfrac{1}{\pi \hbar \sqrt{N}} e^{\frac{-2\sigma^2}{\hbar^2}p^2} \left[ e^{-\frac{(q+\Delta x/2)^2}{2\sigma^2}}+ e^{-\frac{(q-\Delta x/2)^2}{2\sigma^2}} + 2 e^{-\frac{q^2}{2\sigma^2}} \cos\left(\frac{\Delta x\; p}{\hbar}\right)\right].\label{eq:wigsinglep}
\end{equation}
For our two-particle state, the total Wigner function is just the product of single-particle Wigner functions, $W^0(x_1,x_2,p_1,p_2) = W^0_1(x_1,p_1) W^0_2(x_2,p_2)$. The final expression is obtained by substituting in Eq.~\eqref{eq:wigsinglep}, $q=x_1+d/2$ for particle 1 and $q=x_2-d/2$ for particle 2, respectively,
\begin{equation}
\begin{split}
 W^0(x_1,x_2,p_1,p_2)= \left(\dfrac{1}{\pi \hbar\sqrt{N}}\right)^2 e^{-\frac{2\sigma^2}{\hbar^2}(p_1^2+p_2^2)^2}&\left[e^{-\frac{(x_1+\frac{d+\Delta x}{2})^2}{2\sigma^2}}+ e^{-\frac{(x_1+\frac{d-\Delta x}{2})^2}{2\sigma^2}} + 2 e^{-\frac{(x_1+\frac{d}{2})^2}{2\sigma^2}} \cos\left(\frac{\Delta x p}{\hbar}\right)\right]\\
 &\left[e^{-\frac{(x_2-\frac{d-\Delta x}{2})^2}{2\sigma^2}}+ e^{-\frac{(x_2-\frac{d+\Delta x}{2})^2}{2\sigma^2}} + 2 e^{-\frac{(x_2-\frac{d}{2})^2}{2\sigma^2}} \cos\left(\frac{\Delta x p}{\hbar}\right)\right].
\end{split}
\end{equation}
The expansion of this product will result in nine contributions that will depend on different distances,
\begin{equation}
W^0_{\text{TOT}}=\sum_{j=1}^9w_j(x_1,x_2,p_1,p_2),\label{eq:wig9contr}
\end{equation}
omitting the arguments $(x_1,x_2,p_1,p_2)$ for readability. The individual terms are
\begin{equation}
\begin{split}
 & w_1=\dfrac{1}{N}\left(\dfrac{1}{\pi\hbar}\right)^2 e^{-\frac{2\sigma^2}{\hbar^2}(p_1^2+p_2^2)} e^{-\frac{1}{2\sigma^2}(x_1+\frac{d+\Delta x}{2})^2}e^{-\frac{1}{2\sigma^2}(x_2-\frac{d-\Delta x}{2})^2}\\
 & w_2=\dfrac{1}{N}\left(\dfrac{1}{\pi\hbar}\right)^2 e^{-\frac{2\sigma^2}{\hbar^2}(p_1^2+p_2^2)} e^{-\frac{1}{2\sigma^2}(x_1+\frac{d+\Delta x}{2})^2}e^{-\frac{1}{2\sigma^2}(x_2-\frac{d+\Delta x}{2})^2}\\
 & w_3=\dfrac{1}{N}\left(\dfrac{1}{\pi\hbar}\right)^2 e^{-\frac{2\sigma^2}{\hbar^2}(p_1^2+p_2^2)} e^{-\frac{1}{2\sigma^2}(x_1+\frac{d+\Delta x}{2})^2}e^{-\frac{1}{2\sigma^2}(x_2-\frac{d}{2})^2}2\text{cos}\left(\frac{p_2\Delta x}{\hbar}\right)\\
 & w_4=\dfrac{1}{N}\left(\dfrac{1}{\pi\hbar}\right)^2 e^{-\frac{2\sigma^2}{\hbar^2}(p_1^2+p_2^2)} e^{-\frac{1}{2\sigma^2}(x_1+\frac{d-\Delta x}{2})^2}e^{-\frac{1}{2\sigma^2}(x_2-\frac{d-\Delta x}{2})^2}\\ 
 & w_5=\dfrac{1}{N}\left(\dfrac{1}{\pi\hbar}\right)^2 e^{-\frac{2\sigma^2}{\hbar^2}(p_1^2+p_2^2)} e^{-\frac{1}{2\sigma^2}(x_1+\frac{d-\Delta x}{2})^2}e^{-\frac{1}{2\sigma^2}(x_2-\frac{d+\Delta x}{2})^2}\\
 & w_6=\dfrac{1}{N}\left(\dfrac{1}{\pi\hbar}\right)^2 e^{-\frac{2\sigma^2}{\hbar^2}(p_1^2+p_2^2)} e^{-\frac{1}{2\sigma^2}(x_1+\frac{d-\Delta x}{2})^2}e^{-\frac{1}{2\sigma^2}(x_2-\frac{d}{2})^2}2\text{cos}\left(\frac{p_2\Delta x}{\hbar}\right)\\
 & w_7=\dfrac{1}{N}\left(\dfrac{1}{\pi\hbar}\right)^2 e^{-\frac{2\sigma^2}{\hbar^2}(p_1^2+p_2^2)} e^{-\frac{1}{2\sigma^2}(x_1+\frac{d}{2})^2}e^{-\frac{1}{2\sigma^2}(x_2-\frac{d-\Delta x}{2})^2}2\text{cos}\left(\frac{p_1\Delta x}{\hbar}\right)\\
 & w_8=\dfrac{1}{N}\left(\dfrac{1}{\pi\hbar}\right)^2 e^{-\frac{2\sigma^2}{\hbar^2}(p_1^2+p_2^2)} e^{-\frac{1}{2\sigma^2}(x_1+\frac{d}{2})^2}e^{-\frac{1}{2\sigma^2}(x_2-\frac{d+\Delta x}{2})^2}2\text{cos}\left(\frac{p_1\Delta x}{\hbar}\right)\\
 & w_9=\dfrac{1}{N}\left(\dfrac{1}{\pi\hbar}\right)^2 e^{-\frac{2\sigma^2}{\hbar^2}(p_1^2+p_2^2)} e^{-\frac{1}{2\sigma^2}(x_1+\frac{d}{2})^2}e^{-\frac{1}{2\sigma^2}(x_2-\frac{d}{2})^2}2\text{cos}\left(\frac{p_1\Delta x}{\hbar}\right)2\text{cos}\left(\frac{p_2\Delta x}{\hbar}\right)\\
\end{split}\label{eq:wigcontribution}
\end{equation}
All these contributions can be formally expressed as
\begin{equation}
 w_j(x_1,x_2,p_1,p_2) = \frac{4}{N(\pi \hbar)^2}
 e^{-\beta(x_1-x_{0,j})^2-\beta(x_2-y_{0,j})^2-\alpha (p_1^2+p_2^2)} \; \frac{\cos(\gamma_{1,j}\, p_1)}{D_{1,j}} \; \frac{\cos(\gamma_{2,j}\, p_2)}{D_{2,j}},
 \label{eq:wigcontr}
\end{equation}
abbreviating $\alpha = 2\sigma^2/\hbar^2$, $\beta = 1/2\sigma^2$, and introducing the the parameter arrays 
\begin{equation}
 \begin{split}
 [\gamma_{1,j}]_{j=1}^9 &= \left[0,0,0,0,0,0, \frac{\Delta x}{\hbar},\frac{\Delta x}{\hbar},\frac{\Delta x}{\hbar} \right], \qquad [\gamma_{2,j}]_{j=1}^9 =\left[0,0,\frac{\Delta x}{\hbar},0,0,\frac{\Delta x}{\hbar},0,0,\frac{\Delta x}{\hbar} \right]; \\
 [D_{1,j}]_{j=1}^9 &= [2,2,2,2,2,2,1,1,1], \qquad [D_{2,j}]_{j=1}^9 = [2,2,1,2,2,1,2,2,1];\\
[x_{0,j}]_{j=1}^9 &=\left[-\frac{d+\Delta x}{2},-\frac{d+\Delta x}{2},-\frac{d+\Delta x}{2},-\frac{d-\Delta x}{2},-\frac{d-\Delta x}{2},-\frac{d-\Delta x}{2},-\frac{d}{2},-\frac{d}{2},-\frac{d}{2}\right], \\
 [y_{0,j}]_{j=1}^9 &=\left[\frac{d-\Delta x}{2},\frac{d+\Delta x}{2},\frac{d}{2},\frac{d-\Delta x}{2},\frac{d+\Delta x}{2},\frac{d}{2},\frac{d-\Delta x}{2},\frac{d+\Delta x}{2},\frac{d}{2}\right] .
 \end{split}\label{eq:parameters}
\end{equation}

\section{Stepwise linear potential} \label{appendix:stepwisepotential}

To model the classical evolution of the Wigner function in phase space under the gravitational interaction, one could use a stepwise linear potential.
Assuming that the mean distance $d$ between the two interfering particles is greater than the arm separation $\Delta x$ and that the Gaussian wave packets are well localized, $\sigma \ll \Delta x \ll d$, the Wigner function of the two-particle state consists of the non-overlapping contributions \eqref{eq:wigcontribution}. Each contribution $w_j$ corresponds to the two particles being localized around an average distance given by $[\bar{x}_j]_{j=1}^9=\abs{x_{0,j}-y_{0,j}}$. Hence we can safely approximate the gravitational force by opposite constant values $F_j = m^2 G/\bar{x}_j^2$ and $- F_j$ acting on particle 1 and 2, respectively, as resulting from a stepwise linear interaction potential.

The classical time evolution of a particle under a constant force $F_j$ amounts to $x(t) = x(0) + p(0) t/m + F_j t^2/2m$ and $p(t) = p(0) + F_j t$. If we neglect the deflection of the position coordinate and only account for the momentum change---a reasonable approximation for the short experimental time scales considered in Ref.~\cite{bose2017spin}---, the classical time evolution of each Wigner function contribution reads as
\begin{equation}
    w_j(x_1,x_2,p_1,p_2,t) = w_j(x_1,x_2,p_1-F_j t,p_2 + F_j t).
    \label{eq:steplinearforce}
\end{equation}

The problem with this solution of the classical Liouville equation based on stepwise constant forces $F_j$ is that it will generally no longer correspond to a valid state transformation. Indeed, we find that the momentum marginal of the two-particle Wigner function, $W(p_1,p_2,t) = \int W(x_1,x_2,p_1,p_2,t)\dd x_1 \dd x_2 = \sum_j w_j (p_1,p_2,t)$, ceases to be a valid probability distribution of observable momenta as it takes on negative values with growing $t$. Specifically, we get
\begin{equation}
    w_j(p_1,p_2,t) = \frac{4}{N\pi \beta \hbar^2} e^{-\alpha (p_1-F_j t)^2 -\alpha (p_2+F_j t)^2} \; \frac{\cos[\gamma_{1,j} (p_1-F_jt)]}{D_{1,j}} \; \frac{\cos[\gamma_{2,j} (p_2+F_jt )]}{D_{2,j}}.
\end{equation}
The negative regions of these functions get displaced with respect to one another so that they no longer cancel under the $j$-sum. Let us quantify the overall negativity of the momentum marginal by
\begin{equation}
\nu = \frac{1}{2} \left[ \int |W(p_1,p_2,t)|\dd p_1\dd p_2-1 \right]. \label{eq:negMomMarginal}
\end{equation}

Our numerical simulation predicts that a small but noticeable negativity of $ \nu \approx 0.17\,\%$ at $t = 2.5\;\text{s}$ for the parameters used in the main text.
We can alleviate this unphysical behaviour by introducing some amount of noise. 
The simplest noise model consists of momentum diffusion acting on each particle at the diffusion rate $D$. 
In the stepwise linear approximation of the potential, each Wigner function contribution then evolves according to
\begin{equation}
 \partial_t w_j (x_1,x_2,p_1,p_2,t) = F_j \left[\partial_{p_2}-\partial_{p_1} \right]  w_j (x_1,x_2,p_1,p_2,t) + \frac{D}{2} \left[\partial^2_{p_2}+\partial^2_{p_1}\right]  w_j (x_1,x_2,p_1,p_2,t).
\end{equation}
The solution, straightforwardly obtained by Fourier-transforming with respect to the momentum coordinates, is a convolution of the zero-noise solution \eqref{eq:steplinearforce} with a Gaussian of growing width $Dt$,
\begin{equation}
    w_j^D (x_1,x_2,p_1,p_2,t) = \frac{1}{2\pi Dt}\int \dd q_1 \dd q_2 e^{-(q_1^2+q_2^2)/2Dt} w_j(x_1,x_2,p_1-q_1-F_j t,p_2 - q_2 + F_j t).
\end{equation}
Inserting the explicit form \eqref{eq:wigcontr} of the $w_j$ and setting $\widetilde{\alpha} = \alpha/(1+2\alpha Dt)$, we arrive at
\begin{align}
    w_j^D (x_1,x_2,p_1,p_2,t) &= \left[\frac{\beta}{\pi} e^{-\beta(x_1-x_{0,j})^2-\beta(x_2-y_{0,j})^2} \right] \label{eq:wjD}\\
    &\times \left\{ \frac{4\widetilde{\alpha}}{N\pi} e^{-\widetilde{\alpha}(p_1-F_jt)^2-\widetilde{\alpha}(p_2+F_jt)^2 -Dt \widetilde{\alpha} (\gamma_{1,j}^2 + \gamma_{2,j}^2)/2\alpha} \frac{\cos \left[ \frac{\widetilde{\alpha}}{\alpha} \gamma_{1,j} (p_1-F_j t) \right]}{D_{1,j}} \frac{\cos \left[ \frac{\widetilde{\alpha}}{\alpha} \gamma_{2,j} (p_2+F_j t) \right]}{D_{2,j}}\right\} \nonumber .
\end{align}
Here, the square-bracketed expression in the first line represents the position marginal (unaffected by the diffusion) and the second line represents the momentum marginal $w_j^D (p_1,p_2,t)$. Broadly speaking, an overall diffusion $Dt \sim (\hbar/\Delta x)^2$ washes out the cosine fringes in some of the $w_j^D$ and thus renders the Wigner function non-negative, amounting to a classical mixture. On the one hand, this rectifies the negativity of the momentum marginal in the stepwise linear approximation. Specifically, at the proposed interrogation time $t=2.5\,$s, we find that the negativity drops below a numerical threshold value of, say, $\nu < 2\times 10^{-7}$, if $Dt \geq 0.25 (\hbar/\Delta x)^2$. On the other hand, diffusion of this magnitude will also decrease the purity of the two-particle state significantly and more strongly than the purity of the single-particle marginal. 
Given that here, the different Wigner contributions \eqref{eq:wjD} have almost zero overlap due to $\sigma \ll \Delta x,d$, the global purity is very well approximated by
\begin{align}
    &\Gamma^D (t) 
    \approx (2\pi\hbar)^2\sum_{j=1}^9 \int \dd x_1 \dd x_2 \dd p_1 \dd p_2 \, [w^D_j(x_1,x_2,p_1,p_2,t)]^2 = \sum_{j=1}^9 \Gamma^D_{1,j} (t) \Gamma^D_{2,j} (t), \qquad \text{with} \\
    & \Gamma^D_{i,j} = \sqrt{\frac{\beta}{2\pi}}\frac{8\hbar \widetilde{\alpha}}{N D^2_{i,j}} e^{-Dt \widetilde{\alpha} \gamma_{i,j}^2/\alpha} \int \dd p \, e^{-2\widetilde{\alpha} p^2} \cos^2 \left[ \frac{\widetilde{\alpha}}{\alpha} \gamma_{i,j} p \right] 
    = \frac{2 e^{-Dt \widetilde{\alpha} \gamma_{i,j}^2/\alpha}}{N D^2_{i,j} \sqrt{1+2\alpha D t}} \left[ 1 + e^{-\widetilde{\alpha}\gamma^2_{i,j}/2\alpha^2} \right] \approx \frac{e^{-Dt \widetilde{\alpha} \gamma_{i,j}^2/\alpha}}{2 D_{i,j} \sqrt{1+2\alpha D t}} . \nonumber 
\end{align}
In the last step, we set $N\approx 4$ and used that $\gamma_{i,j}^2/\alpha \gg 1$ whenever $\gamma_{i,j}\neq 0$. For the reduced state of particle 2, we must marginalize the Wigner contributions \eqref{eq:wjD} over $x_1,p_1$, which results in
\begin{align}
    w_j^D (x_2,p_2,t) &= \frac{\sqrt{\beta\widetilde\alpha}}{\pi}e^{-\beta(x_2-y_{0,j})^2 -\widetilde{\alpha}(p_2+F_jt)^2 -Dt \widetilde{\alpha} \gamma_{2,j}^2/2\alpha} \frac{\cos \left[ \frac{\widetilde{\alpha}}{\alpha} \gamma_{2,j} (p_2+F_j t) \right]}{2D_{2,j}} \left[ \frac{8}{N D_{1,j}} e^{-\gamma_{1,j}^2/4\alpha} \right] . \label{eq:wjD_particle2}
\end{align}
Here, we can consistently approximate the square-bracketed rightmost expression by one if $\gamma_{1,j}=0$ (i.e.~$j\leq 6$) and zero otherwise. 
Noting that different position displacements $y_{0,j}$ correspond to non-overlapping Gaussians, the purity of the reduced state is very well approximated by
\begin{align}
    \gamma^D (t) &\approx 2\pi\hbar \int\dd x_2 \dd p_2 \sum_{j=1}^3\left[ w_j^D (x_2,p_2,t)+ w_{j+3}^D (x_2,p_2,t) \right]^2 \\
    &= \sqrt{\frac{\beta}{2\pi}} \frac{\hbar \widetilde\alpha}{2} \int\dd p_2 \left\{ \frac{1}{4}\left[ e^{-\widetilde\alpha (p_2+F_1 t)^2} + e^{-\widetilde\alpha (p_2+F_4 t)^2} \right]^2 + \frac{1}{4} \left[ e^{-\widetilde\alpha (p_2+F_2 t)^2} + e^{-\widetilde\alpha (p_2+F_5 t)^2} \right]^2 \right. \nonumber \\
    &\left. \quad + e^{-Dt (\Delta x/\hbar)^2 \widetilde\alpha/\alpha}\left[ e^{-\widetilde\alpha (p_2+F_3 t)^2} \cos \left[ \frac{\widetilde{\alpha}\Delta x}{\alpha \hbar}  (p_2+F_3 t) \right] + e^{-\widetilde\alpha (p_2+F_6 t)^2} \cos \left[ \frac{\widetilde{\alpha}\Delta x}{\alpha \hbar}  (p_2+F_6 t) \right] \right]^2  \right\} \nonumber \\
    &\approx \frac{1}{4}\sum_{j=1}^6 \Gamma_{2,j}^D + \frac{\Gamma_{2,1}^D}{2} \left[ e^{-\widetilde\alpha (F_1-F_4)^2 t^2/2} + e^{-\widetilde\alpha (F_2-F_5)^2 t^2/2} \right] + \frac{\Gamma_{2,3}^D}{2} e^{-\widetilde\alpha (F_3-F_6)^2 t^2/2} \cos \left[ \frac{\widetilde{\alpha}\Delta x}{\alpha \hbar}  (F_3 - F_6) t \right] \nonumber . 
\end{align}
For $\Gamma^D (t) < \gamma^D (t)$, entanglement is no longer detectable (based on the purity criterion). Indeed, for the chosen $Dt$ at $t = 2.5\,$s, we already find $\Gamma_D(t) = 0.79$ and $\gamma^D(t) = 0.88$.

This highlights that it is not at all clear how to construct tractable higher-than-2nd order approximations to the gravitational potential that lead to valid quantum states under the classical Liouville equation and approximate the quantum prediction of gravity-mediated entanglement well. For a more conclusive test of classical models in future proposals, one should therefore explore parameter regimes in which a second-order approximation of the gravitational potential would cease to be valid.

\section{Approximations of the potential} \label{appendix:approx}
In order to find the classical time evolution of the Wigner function we need to solve the Hamilton's equation of motions with the respective approximation of the potential. From the Hamiltonian in the new coordinates given by Eq.~\eqref{eq:newCoordinates} we can expand the Newtonian potential $V_\newton$ in Taylor series in $x_\rel$
\begin{equation}
V_\newton(x_\rel) = - \dfrac{\kappa}{d + \sqrt{2} x_\rel} = \kappa \sum_{n=0}^\infty \dfrac{(-1)^{n+1} 2^{\frac{n}{2}} x_\rel^n}{d^{n+1}} = - \dfrac{\kappa}{d} \sum_{n=0}^\infty \left(\dfrac{- \sqrt{2} x_\rel}{d} \right)^n
\end{equation}
where the expansion converges for $\abs{x_\rel} < \frac{d}{\sqrt{2}}$. After dropping the terms of the order $\order{x_\rel^3}$ and higher we get the potential
\begin{equation}
V_\taylor(x_\rel) =  - \dfrac{\kappa}{d} + \dfrac{\sqrt{2} \kappa}{d^2} x_\rel - \dfrac{2 \kappa}{d^3} x_\rel^2,
\end{equation}
Between two out of the four possible combinations of paths we have $x_\rel = 0$ and so the expansion gives the exact result, in the other two cases we have $x_\rel = \pm \frac{\Delta x}{\sqrt{2}}$ and we get $\frac{\Delta x}{d} \approx 0.556$, where we have used the previously proposed values $\Delta x = 250 \mu \text{m}$ and $d = 450 \mu \text{m}$. It follows that while the Taylor expansion converges, we should not expect to obtain exact results by replacing the Newtonian potential $V_\newton$ with $V_\taylor$.

\begin{figure}
\subfloat[Comparison of the Newtonian potential $V_\newton$ with the Taylor expansion approximation $V_\taylor$ and with the approximation $V_\fit$.\label{fig:potentials}]{\includegraphics[width=0.475\textwidth]{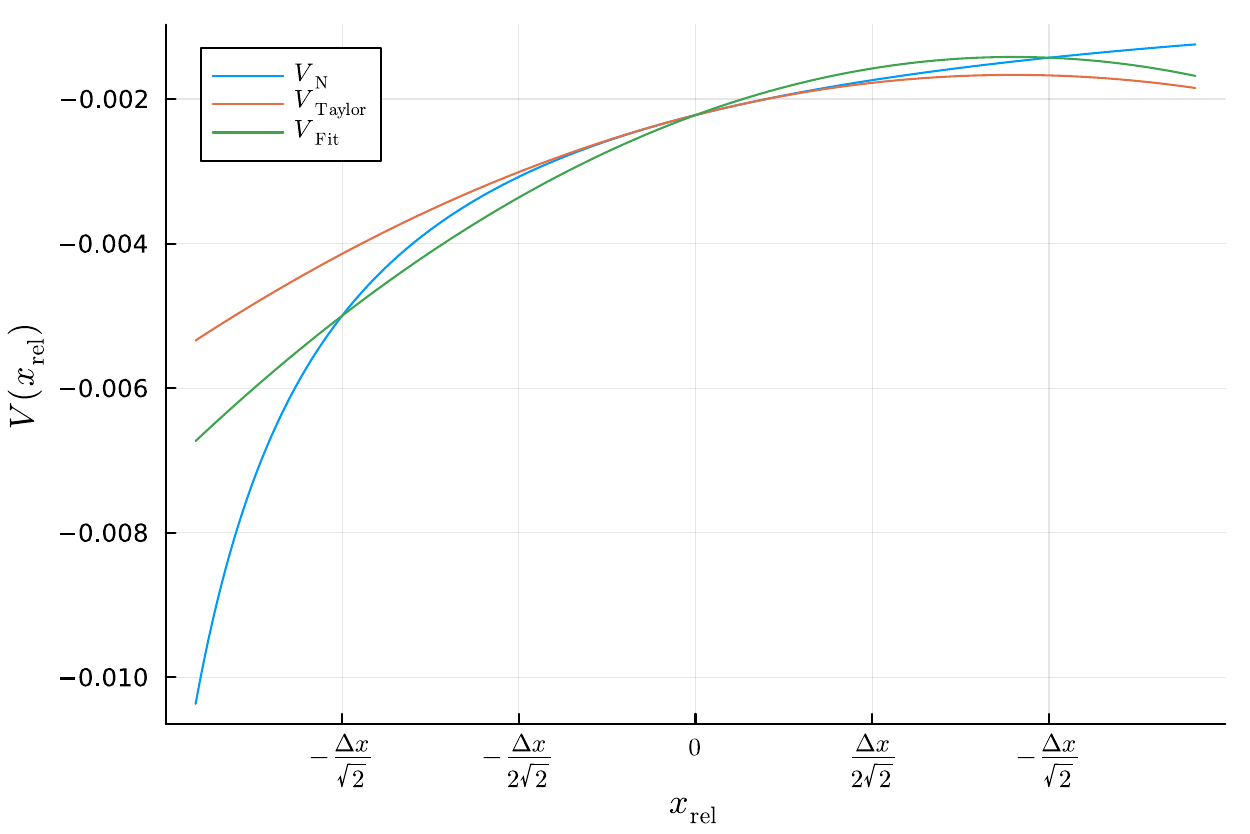}}
\hfill
\subfloat[Comparison of the derivative of the Newtonian potential $V'_\newton$ with the derivatives of the Taylor expansion approximation $V'_\taylor$ and with the approximation $V'_\fit$.\label{fig:potentials-dx}]{\includegraphics[width=0.475\textwidth]{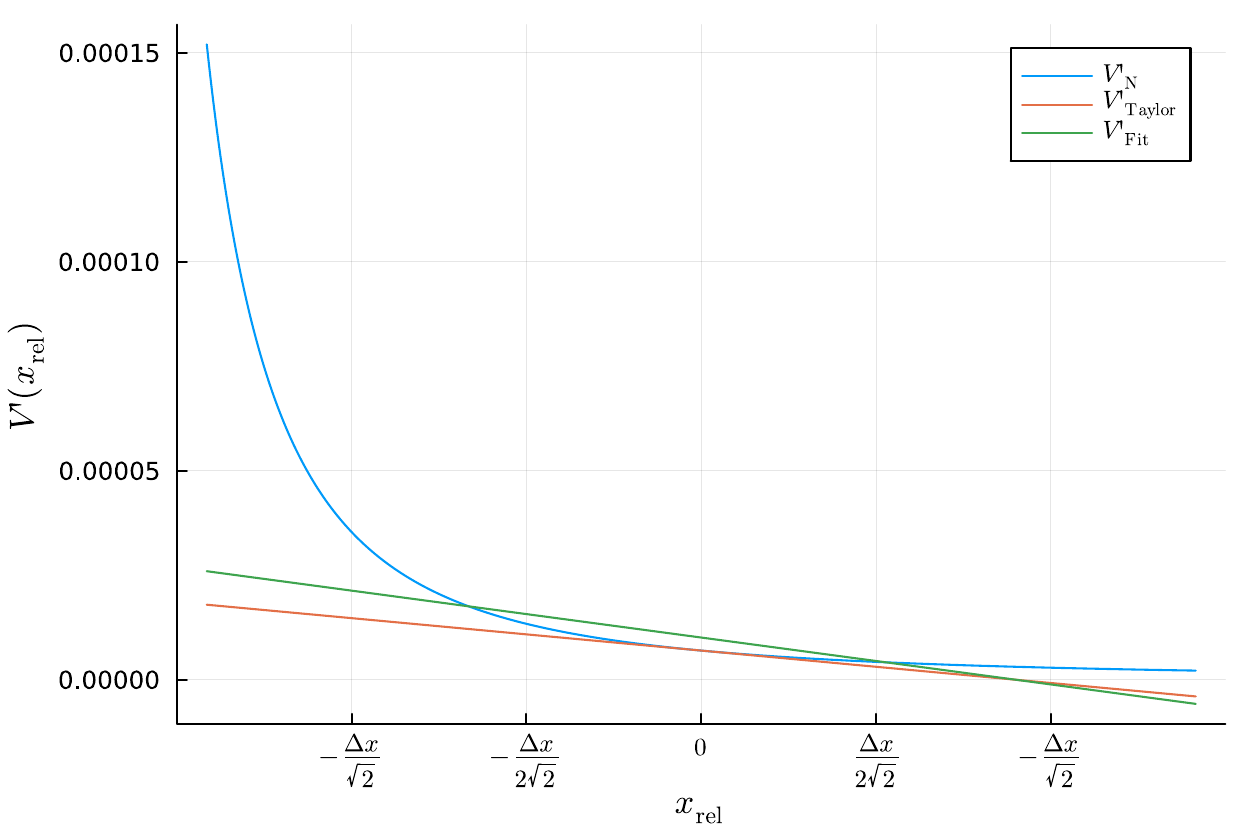}}
\caption{Comparison of the potentials and their derivatives. Previously proposed values $\Delta x = 250 \mu \text{m}$ and $d = 450 \mu \text{m}$ were used; the horizontal axis is in the units of $\Delta x$ while the vertical axis is in the units of $\kappa$.}
\label{fig:potentials-fig}
\end{figure}

The potential $V_\fit$, given as,
\begin{equation}
V_\fit(x_\rel) = - \dfrac{\kappa}{d} + \dfrac{\sqrt{2} \kappa}{d^2 - \Delta x^2} x_\rel - \dfrac{2 \kappa}{d ( d^2 - \Delta x^2 )} x_\rel^2,
\end{equation}
was computed such that it coincides with Newtonian potential for $x_\rel = 0$ and $x_\rel = \pm \frac{\Delta x}{\sqrt{2}}$. Nevertheless one must note that it is not the potential, but the derivative of the potential that appears in the Liouville equation. Let $V'_\taylor$ and $V'_\fit$ denote the derivatives of the respective potential, then one should ask how well $V'_\taylor$ and $V'_\fit$ approximate the derivative of the Newtonian potential $V'_\newton$. The potentials are compared in Fig.~\ref{fig:potentials} and their derivatives are compared in Fig.~\ref{fig:potentials-dx}, one can see that the approximations are reasonably close and that $V'_\fit$ is a better approximation of $V'_\newton$ than $V'_\taylor$.

\section{Classical time evolution} \label{appendix:timeEvo}
Using the potential $V_\taylor(x_\rel)$ we get the Hamiltonian
\begin{equation}
H(x_\avg, x_\rel, p_\avg, p_\rel) = \dfrac{p_\avg^2 + p_\rel^2}{2m} - \dfrac{\kappa}{d} + \dfrac{\sqrt{2} \kappa}{d^2} x_\rel - \dfrac{2 \kappa}{d^3} x_\rel^2
\end{equation}
which yields the equations of motion
\begin{align}
&\dot{x}_\avg = \dfrac{p_\avg}{m},
&&\dot{p}_\avg = 0, \\
&\dot{x}_\rel = \dfrac{p_\rel}{m},
&&\dot{p}_\rel = - \dfrac{\sqrt{2} \kappa}{d^2} + \dfrac{4 \kappa}{d^3} x_\rel. \label{eq:preldot}
\end{align}
We immediately have
\begin{align}
&x_\avg(t) = \dfrac{p_\avg(0)}{m} t + x_\avg(0),
&&p_\avg(t) = p_\avg(0).
\end{align}
To solve the other two equations we compute the second time derivative of $p_\rel$, we get
\begin{equation}
\ddot{p}_\rel = \dfrac{4 \kappa}{d^3} \dot{x}_\rel = \dfrac{4 \kappa}{m d^3} p_\rel
\end{equation}
which is solved by the hyperbolic functions
\begin{equation}
p_\rel(t) = A \sinh \left(2 \sqrt{\dfrac{\kappa}{m d^3}}t\right) + p_\rel(0) \cosh \left(2 \sqrt{\dfrac{\kappa}{m d^3}}t\right)
\end{equation}
where $A$ is a constant to be determined from the initial conditions. We also get
\begin{equation}
x_\rel(t) = \dfrac{A}{2} \sqrt{\dfrac{d^3}{\kappa m}} \cosh \left(2 \sqrt{\dfrac{\kappa}{m d^3}}t\right) + \dfrac{p_\rel(0)}{2} \sqrt{\dfrac{d^3}{\kappa m}} \sinh \left(2 \sqrt{\dfrac{\kappa}{m d^3}}t\right) + B,
\end{equation}
where $B$ is an integration constant. First of all we need to enforce that \eqref{eq:preldot} holds, we have
\begin{equation}
\dot{p}_\rel(t) = 2 A \sqrt{\dfrac{\kappa}{m d^3}} \cosh \left(2 \sqrt{\dfrac{\kappa}{m d^3}} t \right) + 2 p_\rel(0) \sqrt{\dfrac{\kappa}{m d^3}} \cosh \left(2 \sqrt{\dfrac{\kappa}{m d^3}} t \right)
\end{equation}
and
\begin{equation}
- \dfrac{\sqrt{2} \kappa}{d^2} + \dfrac{4 \kappa}{d^3} x_\rel = - \dfrac{\sqrt{2} \kappa}{d^2} + 2 A \sqrt{\dfrac{\kappa}{m d^3}} \cosh \left(2 \sqrt{\dfrac{\kappa}{m d^3}}t\right) + 2 p_\rel(0) \sqrt{\dfrac{\kappa}{m d^3}} \sinh \left(2 \sqrt{\dfrac{\kappa}{m d^3}}t\right) + \dfrac{4 \kappa}{d^3} B,
\end{equation}
By comparing the two expressions we get
\begin{equation}
\dfrac{4 \kappa}{d^3} B = \dfrac{\sqrt{2} \kappa}{d^2}, \qquad \text{i.e.,} \quad B = \dfrac{d}{2 \sqrt{2}}.
\end{equation}
Finally we obtain $A$ from the initial condition
\begin{equation}
x_\rel(0) = \dfrac{A}{2} \sqrt{\dfrac{d^3}{\kappa m}} + \dfrac{d}{2 \sqrt{2}}, \qquad \text{which gives} \quad A = 2 \sqrt{\dfrac{\kappa m}{d^3}} \left( x_\rel(0) - \dfrac{d}{2 \sqrt{2}} \right).
\end{equation}
This leaves us with
\begin{align}
x_\rel(t) &= \left( x_\rel(0) - \dfrac{d}{2 \sqrt{2}} \right) \cosh \left(2 \sqrt{\dfrac{\kappa}{m d^3}}t\right) + \dfrac{p_\rel(0)}{2} \sqrt{\dfrac{d^3}{\kappa m}} \sinh \left(2 \sqrt{\dfrac{\kappa}{m d^3}}t\right) + \dfrac{d}{2 \sqrt{2}}, \\
p_\rel(t) &= 2 \sqrt{\dfrac{\kappa m}{d^3}} \left( x_\rel(0) - \dfrac{d}{2 \sqrt{2}} \right) \sinh \left(2 \sqrt{\dfrac{\kappa}{m d^3}}t\right) + p_\rel(0) \cosh \left(2 \sqrt{\dfrac{\kappa}{m d^3}}t\right).
\end{align}
We obtain the solution in terms of the original coordinates by using the inverse transformation
\begin{align}
&x_1 = \dfrac{1}{\sqrt{2}}(x_\avg - x_\rel - \dfrac{d}{\sqrt{2}}),
&&p_1 = \dfrac{1}{\sqrt{2}}(p_\avg - p_\rel), \\
&x_2 = \dfrac{1}{\sqrt{2}}(x_\avg + x_\rel + \dfrac{d}{\sqrt{2}}),
&&p_2 = \dfrac{1}{\sqrt{2}}(p_\avg + p_\rel).
\end{align}
We arrive at
\begin{align}
x_1(t) &= \dfrac{1}{\sqrt{2}} \left( \dfrac{p_\avg(0)}{m} t + x_\avg(0) - \left( x_\rel(0) - \dfrac{d}{2 \sqrt{2}} \right) \cosh \left(2 \sqrt{\dfrac{\kappa}{m d^3}}t\right) - \dfrac{p_\rel(0)}{2} \sqrt{\dfrac{d^3}{\kappa m}} \sinh \left(2 \sqrt{\dfrac{\kappa}{m d^3}}t\right) - \dfrac{3d}{2 \sqrt{2}} \right),
\\
x_2(t) &= \dfrac{1}{\sqrt{2}} \left( \dfrac{p_\avg(0)}{m} t + x_\avg(0) + \left( x_\rel(0) - \dfrac{d}{2 \sqrt{2}} \right) \cosh \left(2 \sqrt{\dfrac{\kappa}{m d^3}}t\right) + \dfrac{p_\rel(0)}{2} \sqrt{\dfrac{d^3}{\kappa m}} \sinh \left(2 \sqrt{\dfrac{\kappa}{m d^3}}t\right) + \dfrac{3d}{2 \sqrt{2}} \right),
\\
p_1(t) &= \dfrac{1}{\sqrt{2}} \left( p_\avg(0) - 2 \sqrt{\dfrac{\kappa m}{d^3}} \left( x_\rel(0) - \dfrac{d}{2 \sqrt{2}} \right) \sinh \left(2 \sqrt{\dfrac{\kappa}{m d^3}}t\right) - p_\rel(0) \cosh \left(2 \sqrt{\dfrac{\kappa}{m d^3}}t\right) \right),
\\
p_2(t) &= \dfrac{1}{\sqrt{2}} \left( p_\avg(0) + 2 \sqrt{\dfrac{\kappa m}{d^3}} \left( x_\rel(0) - \dfrac{d}{2 \sqrt{2}} \right) \sinh \left(2 \sqrt{\dfrac{\kappa}{m d^3}}t\right) + p_\rel(0) \cosh \left(2 \sqrt{\dfrac{\kappa}{m d^3}}t\right) \right),
\end{align}
which simplifies to
\begin{align}
x_1(t) &= \dfrac{p_1(0) + p_2(0)}{2m} t + \dfrac{x_1(0) + x_2(0)}{2} - \dfrac{x_2(0) - x_1(0) - \frac{3d}{2}}{2} \cosh \left(2 \sqrt{\dfrac{\kappa}{m d^3}}t\right) \nonumber \\
&- \dfrac{p_2(0) - p_1(0)}{4} \sqrt{\dfrac{d^3}{\kappa m}} \sinh \left(2 \sqrt{\dfrac{\kappa}{m d^3}}t\right) - \dfrac{3d}{4},
\\
x_2(t) &= \dfrac{p_1(0) + p_2(0)}{2m} t + \dfrac{x_1(0) + x_2(0)}{2} + \dfrac{x_2(0) - x_1(0) - \frac{3d}{2}}{2} \cosh \left(2 \sqrt{\dfrac{\kappa}{m d^3}}t\right) \nonumber \\
&+ \dfrac{p_2(0) - p_1(0)}{4} \sqrt{\dfrac{d^3}{\kappa m}} \sinh \left(2 \sqrt{\dfrac{\kappa}{m d^3}}t\right) + \dfrac{3d}{4},
\\
p_1(t) &= \dfrac{p_1(0) + p_2(0)}{2} - 2 \sqrt{\dfrac{\kappa m}{d^3}} \dfrac{x_2(0) - x_1(0) - \frac{3d}{2}}{2} \sinh \left(2 \sqrt{\dfrac{\kappa}{m d^3}}t\right) - \dfrac{p_2(0) - p_1(0)}{2} \cosh \left(2 \sqrt{\dfrac{\kappa}{m d^3}}t\right),
\\
p_2(t) &= \dfrac{p_1(0) + p_2(0)}{2} + 2 \sqrt{\dfrac{\kappa m}{d^3}} \dfrac{x_2(0) - x_1(0) - \frac{3d}{2}}{2} \sinh \left(2 \sqrt{\dfrac{\kappa}{m d^3}}t\right) + \dfrac{p_2(0) - p_1(0)}{2} \cosh \left(2 \sqrt{\dfrac{\kappa}{m d^3}}t\right).
\end{align}
where we have also used
\begin{equation}
x_\rel(0) - \dfrac{d}{2 \sqrt{2}} = \dfrac{1}{\sqrt{2}} \left( x_2(0) - x_1(0) - \dfrac{3d}{2} \right).
\end{equation}
In an analogous way, it is possible to derive the equation of motions under the second-order polynomial approximation of the potential $V_\fit$. By using the approximated potential from Eq.~\eqref{eq:Vfit} we obtain the Hamiltonian as
\begin{equation}
H(x_\avg, x_\rel, p_\avg, p_\rel) = \dfrac{p_\avg^2 + p_\rel^2}{2m} - \dfrac{\kappa}{d} + \dfrac{\sqrt{2} \kappa}{d^2-\Delta x^2} x_\rel - \dfrac{2 \kappa}{d(d^2-\Delta x^2)} x_\rel^2.
\end{equation}
Using analogical calculations as in the previous case, we get the solutions to the Hamilton's equations
\begin{align}
x_1(t) &= \dfrac{p_1(0)+p_2(0)}{2m}t+\dfrac{x_1(0)+x_2(0)}{2}-\dfrac{x_2(0)-x_1(0)-\frac{3d}{2}}{2}\cosh \left(2\sqrt{\dfrac{\kappa}{d(d^2-\Delta x^2)m}}t\right) \nonumber \\
&-\dfrac{p_2(0)-p_1(0)}{4}\sqrt{\dfrac{d(d^2-\Delta x^2)}{\kappa m}}\sinh \left(2\sqrt{\dfrac{\kappa}{d(d^2-\Delta x^2)m}}t\right)-\dfrac{3d}{4}\\
x_2(t) &= \dfrac{p_1(0)+p_2(0)}{4}t+\dfrac{x_1(0_+x_2(0)}{2}+\dfrac{x_2(0)-x_1(0)-\frac{3d}{2}}{2}\cosh \left(2\sqrt{\dfrac{\kappa}{d(d^2-\Delta x^2)m}}t\right) \nonumber \\
&+\dfrac{p_2(0)-p_1(0)}{4}\sqrt{\dfrac{d(d^2-\Delta x^2)}{\kappa m}}\sinh \left( 2\sqrt{\dfrac{\kappa}{d(d^2-\Delta x^2)m}}t \right)+\dfrac{3d}{4}\\
p_1(t) &= \dfrac{p_1(0)+p_2(0)}{2}-2\sqrt{\dfrac{\kappa m}{d(d^2-\Delta x^2)}}\dfrac{x_2(0)-x_1(0)-\frac{3d}{2}}{2} \sinh \left( 2\sqrt{\dfrac{\kappa}{d(d^2-\Delta x^2)m}}t \right) \nonumber \\
&-\dfrac{p_2(0)-p_1(0)}{2} \cosh \left(2\sqrt{\dfrac{\kappa }{d(d^2-\Delta x^2)m}}t\right)\\
p_2(t) &= \dfrac{p_1(0)+p_2(0)}{2}+2\sqrt{\dfrac{\kappa m}{d(d^2-\Delta x^2)}}\dfrac{x_2(0)-x_1(0)-\frac{3d}{2}}{2} \sinh \left( 2\sqrt{\dfrac{\kappa}{d(d^2-\Delta x^2)m}}t \right) \nonumber \\
&+\dfrac{p_2(0)-p_1(0)}{2} \cosh \left(2\sqrt{\dfrac{\kappa }{d(d^2-\Delta x^2)m}}t\right)
\end{align}

\section{Comparison of trajectories}
\label{appendix:benchmarktraj}

Finally, we check that our various approximations of the Newtonian potential are valid and match closely the exact classical trajectory for the relative motion of the two particles within the range of evolution times considered. To this end, we compare the approximated classical phase-space trajectories to the solution $[x_{\rel} (t),p_{\rel} (t)]$ of the exact canonical equations of motion based on the Hamiltonian \eqref{eq:Hamiltonian}, 
\begin{equation}
\dot{x}_{\rel} = \frac{1}{m}p_{\rel}, \qquad \dot{p}_{\rel} = - V'_\newton(x_\rel) = - \frac{\sqrt{2}\kappa}{(d+\sqrt{2}x_{\rel})^2} 
\end{equation}
For the two second-order approximations employed in the main text, the Newtonian potential $V_\newton$ in the canonical equation for $p_\rel$ is replaced by the Taylor approximation $V_\taylor$ and the optimized fit $V_\fit$; the respective solutions are given in Appendix \ref{appendix:timeEvo}. Moreover, the stepwise linear approximation in Appendix \ref{appendix:stepwisepotential} amounts to setting $p_\rel (t) \approx p_\rel(0) - \sqrt{2}F_j t$ with $F_j$ chosen according to the initial condition, while omitting any change in position, $x_\rel (t) \approx x_\rel (0)$.

From the nine contributions to the initial Wigner function of the two particles in \eqref{eq:wigcontribution}, we infer that the Wigner function is sensitive to changes in position of the order of $\sigma$ and to changes in momentum of the order of $\hbar/\Delta x$ (for $j=3,6,7,8,9$). Therefore, our approximated trajectories are valid as long as their deviations from the exact trajectory are smaller than those two scales. 

\begin{figure}
    \centering
    \subfloat[Trajectories $x_{rel}(t)$ in units of $\sigma$ for the closest branches of the Wigner function ($j=4$), corresponding to $x_2-x_1=d-\Delta x$. We compare the three different cases: the Newtonian potential $V_N$, the Taylor expansion $V_{Taylor}$, and the polynomial approximation $V_{FIT}$. ]{\includegraphics[width=0.44\textwidth]{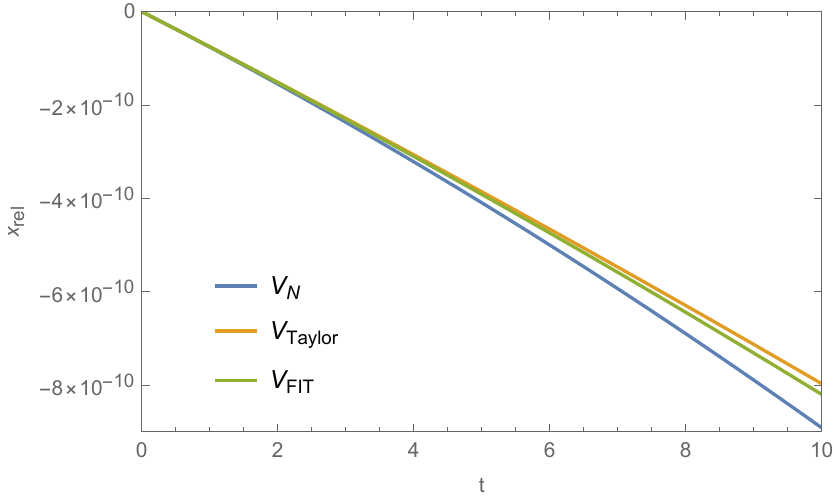}}
    \hfill\subfloat[Comparison of $p_{rel}(t)$ for the branches separated by $x_2-x_1=d-\dfrac{\Delta x}{2}$, corresponding to $j=6,7$ contributions of the Wigner function. We show the four cases: the Newtonian potential $V_N$, the Taylor expansion $V_{Taylor}$, the polynomial approximation $V_{FIT}$, and the step wise potential $V_{Step}$.]{\includegraphics[width=0.4\textwidth]{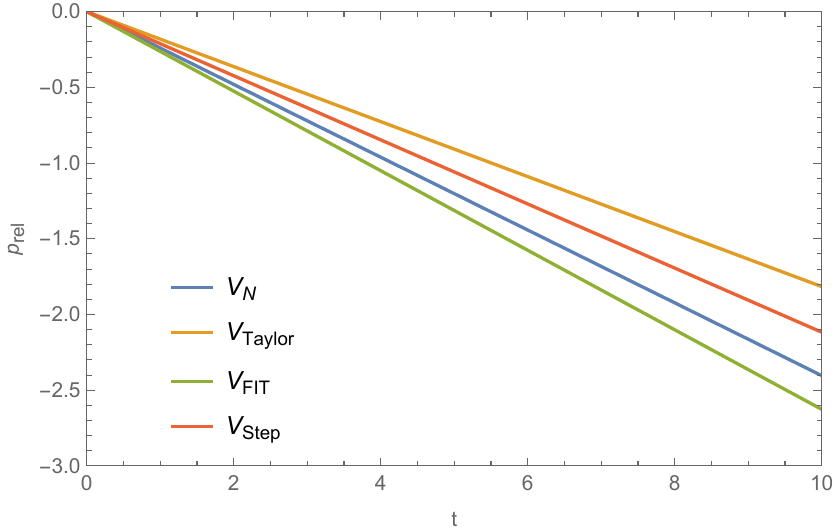}}
    \caption{}
    \label{fig:trajbenchmark}
\end{figure}

Figure \ref{fig:trajbenchmark} compares the three approximations against the exact benchmark in terms of (a) $x_\rel (t) - x_\rel(0)$ in units of $\sigma$ and (b) $p_\rel (t) - p_\rel (0)$ in units of $\hbar/\Delta x$. The plotted time interval matches that of Fig.~\ref{fig:purity} in the main text. In (a), we consider as a worst-case initial condition the branch ($j=4$) in which the two particles are closest (and the attraction the strongest), and we subtract two Gaussian standard deviations in addition, $x_\rel (0) = -(\Delta x + 2\sigma)/\sqrt{2}$ and $p_\rel (0) = - \hbar/\sqrt{2}\sigma$. Since $x_\rel (t)$ barely moves away from $x_\rel (0)$ by $10^{-9}\sigma$, all approximations are almost exact. 
In (b), we choose the worst-case branches ($j=6,7$) with momentum oscillations on the critical scale $\hbar/\Delta x$, setting $x_\rel (0) = -(\Delta x/2 + 2\sigma)/\sqrt{2}$. We find that the optimized quadratic fit and the stepwise linear approximation still match the exact $p_\rel(t)$ fairly well, while the Taylor approximation can deviate more appreciably by up to $0.6\hbar/\Delta x$. 
Judging from these worst-case deviations, we conclude that our approximations are safely valid up to a few times the proposed experimental time scale of $t=2.5\,$s.


\begin{thebibliography}{70}%
\makeatletter
\providecommand \@ifxundefined [1]{%
 \@ifx{#1\undefined}
}%
\providecommand \@ifnum [1]{%
 \ifnum #1\expandafter \@firstoftwo
 \else \expandafter \@secondoftwo
 \fi
}%
\providecommand \@ifx [1]{%
 \ifx #1\expandafter \@firstoftwo
 \else \expandafter \@secondoftwo
 \fi
}%
\providecommand \natexlab [1]{#1}%
\providecommand \enquote  [1]{``#1''}%
\providecommand \bibnamefont  [1]{#1}%
\providecommand \bibfnamefont [1]{#1}%
\providecommand \citenamefont [1]{#1}%
\providecommand \href@noop [0]{\@secondoftwo}%
\providecommand \href [0]{\begingroup \@sanitize@url \@href}%
\providecommand \@href[1]{\@@startlink{#1}\@@href}%
\providecommand \@@href[1]{\endgroup#1\@@endlink}%
\providecommand \@sanitize@url [0]{\catcode `\\12\catcode `\$12\catcode
  `\&12\catcode `\#12\catcode `\^12\catcode `\_12\catcode `\%12\relax}%
\providecommand \@@startlink[1]{}%
\providecommand \@@endlink[0]{}%
\providecommand \url  [0]{\begingroup\@sanitize@url \@url }%
\providecommand \@url [1]{\endgroup\@href {#1}{\urlprefix }}%
\providecommand \urlprefix  [0]{URL }%
\providecommand \Eprint [0]{\href }%
\providecommand \doibase [0]{https://doi.org/}%
\providecommand \selectlanguage [0]{\@gobble}%
\providecommand \bibinfo  [0]{\@secondoftwo}%
\providecommand \bibfield  [0]{\@secondoftwo}%
\providecommand \translation [1]{[#1]}%
\providecommand \BibitemOpen [0]{}%
\providecommand \bibitemStop [0]{}%
\providecommand \bibitemNoStop [0]{.\EOS\space}%
\providecommand \EOS [0]{\spacefactor3000\relax}%
\providecommand \BibitemShut  [1]{\csname bibitem#1\endcsname}%
\let\auto@bib@innerbib\@empty
\bibitem [{\citenamefont {Bose}\ \emph {et~al.}(2017)\citenamefont {Bose},
  \citenamefont {Mazumdar}, \citenamefont {Morley}, \citenamefont {Ulbricht},
  \citenamefont {Toro{\v{s}}}, \citenamefont {Paternostro}, \citenamefont
  {Geraci}, \citenamefont {Barker}, \citenamefont {Kim},\ and\ \citenamefont
  {Milburn}}]{bose2017spin}%
  \BibitemOpen
  \bibfield  {author} {\bibinfo {author} {\bibfnamefont {S.}~\bibnamefont
  {Bose}}, \bibinfo {author} {\bibfnamefont {A.}~\bibnamefont {Mazumdar}},
  \bibinfo {author} {\bibfnamefont {G.~W.}\ \bibnamefont {Morley}}, \bibinfo
  {author} {\bibfnamefont {H.}~\bibnamefont {Ulbricht}}, \bibinfo {author}
  {\bibfnamefont {M.}~\bibnamefont {Toro{\v{s}}}}, \bibinfo {author}
  {\bibfnamefont {M.}~\bibnamefont {Paternostro}}, \bibinfo {author}
  {\bibfnamefont {A.~A.}\ \bibnamefont {Geraci}}, \bibinfo {author}
  {\bibfnamefont {P.~F.}\ \bibnamefont {Barker}}, \bibinfo {author}
  {\bibfnamefont {M.}~\bibnamefont {Kim}},\ and\ \bibinfo {author}
  {\bibfnamefont {G.}~\bibnamefont {Milburn}},\ }\bibfield  {title} {\bibinfo
  {title} {Spin entanglement witness for quantum gravity},\ }\href@noop {}
  {\bibfield  {journal} {\bibinfo  {journal} {Physical review letters}\
  }\textbf {\bibinfo {volume} {119}},\ \bibinfo {pages} {240401} (\bibinfo
  {year} {2017})}\BibitemShut {NoStop}%
\bibitem [{\citenamefont {Marletto}\ and\ \citenamefont
  {Vedral}(2017)}]{marletto2017gravitationally}%
  \BibitemOpen
  \bibfield  {author} {\bibinfo {author} {\bibfnamefont {C.}~\bibnamefont
  {Marletto}}\ and\ \bibinfo {author} {\bibfnamefont {V.}~\bibnamefont
  {Vedral}},\ }\bibfield  {title} {\bibinfo {title} {Gravitationally induced
  entanglement between two massive particles is sufficient evidence of quantum
  effects in gravity},\ }\href@noop {} {\bibfield  {journal} {\bibinfo
  {journal} {Physical review letters}\ }\textbf {\bibinfo {volume} {119}},\
  \bibinfo {pages} {240402} (\bibinfo {year} {2017})}\BibitemShut {NoStop}%
\bibitem [{\citenamefont {Al~Balushi}\ \emph {et~al.}(2018)\citenamefont
  {Al~Balushi}, \citenamefont {Cong},\ and\ \citenamefont
  {Mann}}]{al2018optomechanical}%
  \BibitemOpen
  \bibfield  {author} {\bibinfo {author} {\bibfnamefont {A.}~\bibnamefont
  {Al~Balushi}}, \bibinfo {author} {\bibfnamefont {W.}~\bibnamefont {Cong}},\
  and\ \bibinfo {author} {\bibfnamefont {R.~B.}\ \bibnamefont {Mann}},\
  }\bibfield  {title} {\bibinfo {title} {Optomechanical quantum cavendish
  experiment},\ }\href@noop {} {\bibfield  {journal} {\bibinfo  {journal}
  {Physical Review A}\ }\textbf {\bibinfo {volume} {98}},\ \bibinfo {pages}
  {043811} (\bibinfo {year} {2018})}\BibitemShut {NoStop}%
\bibitem [{\citenamefont {Krisnanda}\ \emph {et~al.}(2020)\citenamefont
  {Krisnanda}, \citenamefont {Tham}, \citenamefont {Paternostro},\ and\
  \citenamefont {Paterek}}]{krisnanda2020observable}%
  \BibitemOpen
  \bibfield  {author} {\bibinfo {author} {\bibfnamefont {T.}~\bibnamefont
  {Krisnanda}}, \bibinfo {author} {\bibfnamefont {G.~Y.}\ \bibnamefont {Tham}},
  \bibinfo {author} {\bibfnamefont {M.}~\bibnamefont {Paternostro}},\ and\
  \bibinfo {author} {\bibfnamefont {T.}~\bibnamefont {Paterek}},\ }\bibfield
  {title} {\bibinfo {title} {Observable quantum entanglement due to gravity},\
  }\href@noop {} {\bibfield  {journal} {\bibinfo  {journal} {npj Quantum
  Information}\ }\textbf {\bibinfo {volume} {6}},\ \bibinfo {pages} {12}
  (\bibinfo {year} {2020})}\BibitemShut {NoStop}%
\bibitem [{\citenamefont {Weiss}\ \emph {et~al.}(2021)\citenamefont {Weiss},
  \citenamefont {Roda-Llordes}, \citenamefont {Torrontegui}, \citenamefont
  {Aspelmeyer},\ and\ \citenamefont {Romero-Isart}}]{weiss2021large}%
  \BibitemOpen
  \bibfield  {author} {\bibinfo {author} {\bibfnamefont {T.}~\bibnamefont
  {Weiss}}, \bibinfo {author} {\bibfnamefont {M.}~\bibnamefont {Roda-Llordes}},
  \bibinfo {author} {\bibfnamefont {E.}~\bibnamefont {Torrontegui}}, \bibinfo
  {author} {\bibfnamefont {M.}~\bibnamefont {Aspelmeyer}},\ and\ \bibinfo
  {author} {\bibfnamefont {O.}~\bibnamefont {Romero-Isart}},\ }\bibfield
  {title} {\bibinfo {title} {Large quantum delocalization of a levitated
  nanoparticle using optimal control: Applications for force sensing and
  entangling via weak forces},\ }\href@noop {} {\bibfield  {journal} {\bibinfo
  {journal} {Physical Review Letters}\ }\textbf {\bibinfo {volume} {127}},\
  \bibinfo {pages} {023601} (\bibinfo {year} {2021})}\BibitemShut {NoStop}%
\bibitem [{\citenamefont {Horodecki}\ \emph {et~al.}(2009)\citenamefont
  {Horodecki}, \citenamefont {Horodecki}, \citenamefont {Horodecki},\ and\
  \citenamefont {Horodecki}}]{horodecki2009quantum}%
  \BibitemOpen
  \bibfield  {author} {\bibinfo {author} {\bibfnamefont {R.}~\bibnamefont
  {Horodecki}}, \bibinfo {author} {\bibfnamefont {P.}~\bibnamefont
  {Horodecki}}, \bibinfo {author} {\bibfnamefont {M.}~\bibnamefont
  {Horodecki}},\ and\ \bibinfo {author} {\bibfnamefont {K.}~\bibnamefont
  {Horodecki}},\ }\bibfield  {title} {\bibinfo {title} {Quantum entanglement},\
  }\href@noop {} {\bibfield  {journal} {\bibinfo  {journal} {Reviews of modern
  physics}\ }\textbf {\bibinfo {volume} {81}},\ \bibinfo {pages} {865}
  (\bibinfo {year} {2009})}\BibitemShut {NoStop}%
\bibitem [{\citenamefont {G{\"u}hne}\ and\ \citenamefont
  {T{\'o}th}(2009)}]{guhne2009entanglement}%
  \BibitemOpen
  \bibfield  {author} {\bibinfo {author} {\bibfnamefont {O.}~\bibnamefont
  {G{\"u}hne}}\ and\ \bibinfo {author} {\bibfnamefont {G.}~\bibnamefont
  {T{\'o}th}},\ }\bibfield  {title} {\bibinfo {title} {Entanglement
  detection},\ }\href@noop {} {\bibfield  {journal} {\bibinfo  {journal}
  {Physics Reports}\ }\textbf {\bibinfo {volume} {474}},\ \bibinfo {pages} {1}
  (\bibinfo {year} {2009})}\BibitemShut {NoStop}%
\bibitem [{\citenamefont {Galley}\ \emph {et~al.}(2022)\citenamefont {Galley},
  \citenamefont {Giacomini},\ and\ \citenamefont {Selby}}]{galley2022no}%
  \BibitemOpen
  \bibfield  {author} {\bibinfo {author} {\bibfnamefont {T.~D.}\ \bibnamefont
  {Galley}}, \bibinfo {author} {\bibfnamefont {F.}~\bibnamefont {Giacomini}},\
  and\ \bibinfo {author} {\bibfnamefont {J.~H.}\ \bibnamefont {Selby}},\
  }\bibfield  {title} {\bibinfo {title} {A no-go theorem on the nature of the
  gravitational field beyond quantum theory},\ }\href@noop {} {\bibfield
  {journal} {\bibinfo  {journal} {Quantum}\ }\textbf {\bibinfo {volume} {6}},\
  \bibinfo {pages} {779} (\bibinfo {year} {2022})}\BibitemShut {NoStop}%
\bibitem [{\citenamefont {Marshman}\ \emph {et~al.}(2020)\citenamefont
  {Marshman}, \citenamefont {Mazumdar},\ and\ \citenamefont
  {Bose}}]{marshman2020locality}%
  \BibitemOpen
  \bibfield  {author} {\bibinfo {author} {\bibfnamefont {R.~J.}\ \bibnamefont
  {Marshman}}, \bibinfo {author} {\bibfnamefont {A.}~\bibnamefont {Mazumdar}},\
  and\ \bibinfo {author} {\bibfnamefont {S.}~\bibnamefont {Bose}},\ }\bibfield
  {title} {\bibinfo {title} {Locality and entanglement in table-top testing of
  the quantum nature of linearized gravity},\ }\href@noop {} {\bibfield
  {journal} {\bibinfo  {journal} {Physical Review A}\ }\textbf {\bibinfo
  {volume} {101}},\ \bibinfo {pages} {052110} (\bibinfo {year}
  {2020})}\BibitemShut {NoStop}%
\bibitem [{\citenamefont {Tilloy}(2019)}]{tilloy2019does}%
  \BibitemOpen
  \bibfield  {author} {\bibinfo {author} {\bibfnamefont {A.}~\bibnamefont
  {Tilloy}},\ }\bibfield  {title} {\bibinfo {title} {Does gravity have to be
  quantized? lessons from non-relativistic toy models},\ }in\ \href@noop {}
  {\emph {\bibinfo {booktitle} {Journal of Physics: Conference Series}}},\
  Vol.\ \bibinfo {volume} {1275}\ (\bibinfo {organization} {IOP Publishing},\
  \bibinfo {year} {2019})\ p.\ \bibinfo {pages} {012006}\BibitemShut {NoStop}%
\bibitem [{\citenamefont {Oppenheim}(2023{\natexlab{a}})}]{oppenheim2023time}%
  \BibitemOpen
  \bibfield  {author} {\bibinfo {author} {\bibfnamefont {J.}~\bibnamefont
  {Oppenheim}},\ }\bibfield  {title} {\bibinfo {title} {Is it time to rethink
  quantum gravity?}} (\bibinfo {year} {2023}{\natexlab{a}}),\ \bibinfo {note}
  {arXiv 2310.12221}\BibitemShut {NoStop}%
\bibitem [{\citenamefont {Aspelmeyer}(2022)}]{aspelmeyer2022zeh}%
  \BibitemOpen
  \bibfield  {author} {\bibinfo {author} {\bibfnamefont {M.}~\bibnamefont
  {Aspelmeyer}},\ }\bibfield  {title} {\bibinfo {title} {When zeh meets
  feynman: How to avoid the appearance of a classical world in gravity
  experiments},\ }in\ \href@noop {} {\emph {\bibinfo {booktitle} {From Quantum
  to Classical: Essays in Honour of H.-Dieter Zeh}}}\ (\bibinfo  {publisher}
  {Springer},\ \bibinfo {year} {2022})\ pp.\ \bibinfo {pages}
  {85--95}\BibitemShut {NoStop}%
\bibitem [{\citenamefont {Nguyen}\ and\ \citenamefont
  {Bernards}(2020)}]{nguyen2020entanglement}%
  \BibitemOpen
  \bibfield  {author} {\bibinfo {author} {\bibfnamefont {H.~C.}\ \bibnamefont
  {Nguyen}}\ and\ \bibinfo {author} {\bibfnamefont {F.}~\bibnamefont
  {Bernards}},\ }\bibfield  {title} {\bibinfo {title} {Entanglement dynamics of
  two mesoscopic objects with gravitational interaction},\ }\href@noop {}
  {\bibfield  {journal} {\bibinfo  {journal} {The European Physical Journal D}\
  }\textbf {\bibinfo {volume} {74}},\ \bibinfo {pages} {1} (\bibinfo {year}
  {2020})}\BibitemShut {NoStop}%
\bibitem [{\citenamefont {Lami}\ \emph {et~al.}(2023)\citenamefont {Lami},
  \citenamefont {Pedernales},\ and\ \citenamefont {Plenio}}]{lami2023testing}%
  \BibitemOpen
  \bibfield  {author} {\bibinfo {author} {\bibfnamefont {L.}~\bibnamefont
  {Lami}}, \bibinfo {author} {\bibfnamefont {J.~S.}\ \bibnamefont
  {Pedernales}},\ and\ \bibinfo {author} {\bibfnamefont {M.~B.}\ \bibnamefont
  {Plenio}},\ }\bibfield  {title} {\bibinfo {title} {Testing the quantumness of
  gravity without entanglement}} (\bibinfo {year} {2023}),\ \bibinfo {note}
  {arXiv 2302.03075}\BibitemShut {NoStop}%
\bibitem [{\citenamefont {Howl}\ \emph {et~al.}(2023)\citenamefont {Howl},
  \citenamefont {Cooper},\ and\ \citenamefont
  {Hackerm{\"u}ller}}]{howl2023gravitationally}%
  \BibitemOpen
  \bibfield  {author} {\bibinfo {author} {\bibfnamefont {R.}~\bibnamefont
  {Howl}}, \bibinfo {author} {\bibfnamefont {N.}~\bibnamefont {Cooper}},\ and\
  \bibinfo {author} {\bibfnamefont {L.}~\bibnamefont {Hackerm{\"u}ller}},\
  }\bibfield  {title} {\bibinfo {title} {Gravitationally-induced entanglement
  in cold atoms}} (\bibinfo {year} {2023}),\ \bibinfo {note} {arXiv
  2304.00734}\BibitemShut {NoStop}%
\bibitem [{\citenamefont {Hanif}\ \emph {et~al.}(2023)\citenamefont {Hanif},
  \citenamefont {Das}, \citenamefont {Halliwell}, \citenamefont {Home},
  \citenamefont {Mazumdar}, \citenamefont {Ulbricht},\ and\ \citenamefont
  {Bose}}]{hanif2023testing}%
  \BibitemOpen
  \bibfield  {author} {\bibinfo {author} {\bibfnamefont {F.}~\bibnamefont
  {Hanif}}, \bibinfo {author} {\bibfnamefont {D.}~\bibnamefont {Das}}, \bibinfo
  {author} {\bibfnamefont {J.}~\bibnamefont {Halliwell}}, \bibinfo {author}
  {\bibfnamefont {D.}~\bibnamefont {Home}}, \bibinfo {author} {\bibfnamefont
  {A.}~\bibnamefont {Mazumdar}}, \bibinfo {author} {\bibfnamefont
  {H.}~\bibnamefont {Ulbricht}},\ and\ \bibinfo {author} {\bibfnamefont
  {S.}~\bibnamefont {Bose}},\ }\bibfield  {title} {\bibinfo {title} {Testing
  whether gravity acts as a quantum entity when measured}} (\bibinfo {year}
  {2023}),\ \bibinfo {note} {arXiv 2307.08133}\BibitemShut {NoStop}%
\bibitem [{\citenamefont {Pesci}\ and\ \citenamefont
  {Pieri}(2023)}]{pesci2023testing}%
  \BibitemOpen
  \bibfield  {author} {\bibinfo {author} {\bibfnamefont {A.}~\bibnamefont
  {Pesci}}\ and\ \bibinfo {author} {\bibfnamefont {P.}~\bibnamefont {Pieri}},\
  }\bibfield  {title} {\bibinfo {title} {Testing the nonclassicality of gravity
  with the field of a single delocalized mass},\ }\href@noop {} {\bibfield
  {journal} {\bibinfo  {journal} {Physical Review A}\ }\textbf {\bibinfo
  {volume} {108}},\ \bibinfo {pages} {062801} (\bibinfo {year}
  {2023})}\BibitemShut {NoStop}%
\bibitem [{\citenamefont {Pedernales}\ \emph {et~al.}(2022)\citenamefont
  {Pedernales}, \citenamefont {Streltsov},\ and\ \citenamefont
  {Plenio}}]{pedernales2022enhancing}%
  \BibitemOpen
  \bibfield  {author} {\bibinfo {author} {\bibfnamefont {J.~S.}\ \bibnamefont
  {Pedernales}}, \bibinfo {author} {\bibfnamefont {K.}~\bibnamefont
  {Streltsov}},\ and\ \bibinfo {author} {\bibfnamefont {M.~B.}\ \bibnamefont
  {Plenio}},\ }\bibfield  {title} {\bibinfo {title} {Enhancing gravitational
  interaction between quantum systems by a massive mediator},\ }\href@noop {}
  {\bibfield  {journal} {\bibinfo  {journal} {Physical Review Letters}\
  }\textbf {\bibinfo {volume} {128}},\ \bibinfo {pages} {110401} (\bibinfo
  {year} {2022})}\BibitemShut {NoStop}%
\bibitem [{\citenamefont {Das}\ \emph {et~al.}(2024)\citenamefont {Das},
  \citenamefont {Home}, \citenamefont {Ulbricht},\ and\ \citenamefont
  {Bose}}]{das2024mass}%
  \BibitemOpen
  \bibfield  {author} {\bibinfo {author} {\bibfnamefont {D.}~\bibnamefont
  {Das}}, \bibinfo {author} {\bibfnamefont {D.}~\bibnamefont {Home}}, \bibinfo
  {author} {\bibfnamefont {H.}~\bibnamefont {Ulbricht}},\ and\ \bibinfo
  {author} {\bibfnamefont {S.}~\bibnamefont {Bose}},\ }\bibfield  {title}
  {\bibinfo {title} {Mass-independent scheme to test the quantumness of a
  massive object},\ }\href@noop {} {\bibfield  {journal} {\bibinfo  {journal}
  {Physical Review Letters}\ }\textbf {\bibinfo {volume} {132}},\ \bibinfo
  {pages} {030202} (\bibinfo {year} {2024})}\BibitemShut {NoStop}%
\bibitem [{\citenamefont {Christodoulou}\ \emph
  {et~al.}(2023{\natexlab{a}})\citenamefont {Christodoulou}, \citenamefont
  {Di~Biagio}, \citenamefont {Howl},\ and\ \citenamefont
  {Rovelli}}]{christodoulou2023gravity}%
  \BibitemOpen
  \bibfield  {author} {\bibinfo {author} {\bibfnamefont {M.}~\bibnamefont
  {Christodoulou}}, \bibinfo {author} {\bibfnamefont {A.}~\bibnamefont
  {Di~Biagio}}, \bibinfo {author} {\bibfnamefont {R.}~\bibnamefont {Howl}},\
  and\ \bibinfo {author} {\bibfnamefont {C.}~\bibnamefont {Rovelli}},\
  }\bibfield  {title} {\bibinfo {title} {Gravity entanglement, quantum
  reference systems, degrees of freedom},\ }\href@noop {} {\bibfield  {journal}
  {\bibinfo  {journal} {Classical and Quantum Gravity}\ }\textbf {\bibinfo
  {volume} {40}},\ \bibinfo {pages} {047001} (\bibinfo {year}
  {2023}{\natexlab{a}})}\BibitemShut {NoStop}%
\bibitem [{\citenamefont {Overstreet}\ \emph {et~al.}(2023)\citenamefont
  {Overstreet}, \citenamefont {Curti}, \citenamefont {Kim}, \citenamefont
  {Asenbaum}, \citenamefont {Kasevich},\ and\ \citenamefont
  {Giacomini}}]{overstreet2023inference}%
  \BibitemOpen
  \bibfield  {author} {\bibinfo {author} {\bibfnamefont {C.}~\bibnamefont
  {Overstreet}}, \bibinfo {author} {\bibfnamefont {J.}~\bibnamefont {Curti}},
  \bibinfo {author} {\bibfnamefont {M.}~\bibnamefont {Kim}}, \bibinfo {author}
  {\bibfnamefont {P.}~\bibnamefont {Asenbaum}}, \bibinfo {author}
  {\bibfnamefont {M.~A.}\ \bibnamefont {Kasevich}},\ and\ \bibinfo {author}
  {\bibfnamefont {F.}~\bibnamefont {Giacomini}},\ }\bibfield  {title} {\bibinfo
  {title} {Inference of gravitational field superposition from quantum
  measurements},\ }\href@noop {} {\bibfield  {journal} {\bibinfo  {journal}
  {Physical Review D}\ }\textbf {\bibinfo {volume} {108}},\ \bibinfo {pages}
  {084038} (\bibinfo {year} {2023})}\BibitemShut {NoStop}%
\bibitem [{\citenamefont {Bengyat}\ \emph {et~al.}(2023)\citenamefont
  {Bengyat}, \citenamefont {Di~Biagio}, \citenamefont {Aspelmeyer},\ and\
  \citenamefont {Christodoulou}}]{bengyat2023gravity}%
  \BibitemOpen
  \bibfield  {author} {\bibinfo {author} {\bibfnamefont {O.}~\bibnamefont
  {Bengyat}}, \bibinfo {author} {\bibfnamefont {A.}~\bibnamefont {Di~Biagio}},
  \bibinfo {author} {\bibfnamefont {M.}~\bibnamefont {Aspelmeyer}},\ and\
  \bibinfo {author} {\bibfnamefont {M.}~\bibnamefont {Christodoulou}},\
  }\bibfield  {title} {\bibinfo {title} {Gravity mediated entanglement between
  oscillators as quantum superposition of geometries},\ }\Eprint
  {https://arxiv.org/abs/2309.16312} {arXiv:2309.16312}  (\bibinfo {year}
  {2023})\BibitemShut {NoStop}%
\bibitem [{\citenamefont {Hilweg}\ \emph {et~al.}(2017)\citenamefont {Hilweg},
  \citenamefont {Massa}, \citenamefont {Martynov}, \citenamefont {Mavalvala},
  \citenamefont {Chru{\'s}ciel},\ and\ \citenamefont
  {Walther}}]{hilweg2017gravitationally}%
  \BibitemOpen
  \bibfield  {author} {\bibinfo {author} {\bibfnamefont {C.}~\bibnamefont
  {Hilweg}}, \bibinfo {author} {\bibfnamefont {F.}~\bibnamefont {Massa}},
  \bibinfo {author} {\bibfnamefont {D.}~\bibnamefont {Martynov}}, \bibinfo
  {author} {\bibfnamefont {N.}~\bibnamefont {Mavalvala}}, \bibinfo {author}
  {\bibfnamefont {P.~T.}\ \bibnamefont {Chru{\'s}ciel}},\ and\ \bibinfo
  {author} {\bibfnamefont {P.}~\bibnamefont {Walther}},\ }\bibfield  {title}
  {\bibinfo {title} {Gravitationally induced phase shift on a single photon},\
  }\href@noop {} {\bibfield  {journal} {\bibinfo  {journal} {New Journal of
  Physics}\ }\textbf {\bibinfo {volume} {19}},\ \bibinfo {pages} {033028}
  (\bibinfo {year} {2017})}\BibitemShut {NoStop}%
\bibitem [{\citenamefont {Qvarfort}\ \emph {et~al.}(2022)\citenamefont
  {Qvarfort}, \citenamefont {R{\"a}tzel},\ and\ \citenamefont
  {Stopyra}}]{qvarfort2022constraining}%
  \BibitemOpen
  \bibfield  {author} {\bibinfo {author} {\bibfnamefont {S.}~\bibnamefont
  {Qvarfort}}, \bibinfo {author} {\bibfnamefont {D.}~\bibnamefont
  {R{\"a}tzel}},\ and\ \bibinfo {author} {\bibfnamefont {S.}~\bibnamefont
  {Stopyra}},\ }\bibfield  {title} {\bibinfo {title} {Constraining modified
  gravity with quantum optomechanics},\ }\href@noop {} {\bibfield  {journal}
  {\bibinfo  {journal} {New Journal of Physics}\ }\textbf {\bibinfo {volume}
  {24}},\ \bibinfo {pages} {033009} (\bibinfo {year} {2022})}\BibitemShut
  {NoStop}%
\bibitem [{\citenamefont {Panda}\ \emph {et~al.}(2023)\citenamefont {Panda},
  \citenamefont {Tao}, \citenamefont {Ceja},\ and\ \citenamefont
  {M{\"u}ller}}]{panda2023measuring}%
  \BibitemOpen
  \bibfield  {author} {\bibinfo {author} {\bibfnamefont {C.~D.}\ \bibnamefont
  {Panda}}, \bibinfo {author} {\bibfnamefont {M.~J.}\ \bibnamefont {Tao}},
  \bibinfo {author} {\bibfnamefont {M.}~\bibnamefont {Ceja}},\ and\ \bibinfo
  {author} {\bibfnamefont {H.}~\bibnamefont {M{\"u}ller}},\ }\bibfield  {title}
  {\bibinfo {title} {Measuring gravity by holding atoms},\ }\Eprint
  {https://arxiv.org/abs/2310.01344} {arXiv:2310.01344}  (\bibinfo {year}
  {2023})\BibitemShut {NoStop}%
\bibitem [{\citenamefont {Danielson}\ \emph {et~al.}(2022)\citenamefont
  {Danielson}, \citenamefont {Satishchandran},\ and\ \citenamefont
  {Wald}}]{danielson2022gravitationally}%
  \BibitemOpen
  \bibfield  {author} {\bibinfo {author} {\bibfnamefont {D.~L.}\ \bibnamefont
  {Danielson}}, \bibinfo {author} {\bibfnamefont {G.}~\bibnamefont
  {Satishchandran}},\ and\ \bibinfo {author} {\bibfnamefont {R.~M.}\
  \bibnamefont {Wald}},\ }\bibfield  {title} {\bibinfo {title} {Gravitationally
  mediated entanglement: Newtonian field versus gravitons},\ }\href@noop {}
  {\bibfield  {journal} {\bibinfo  {journal} {Physical Review D}\ }\textbf
  {\bibinfo {volume} {105}},\ \bibinfo {pages} {086001} (\bibinfo {year}
  {2022})}\BibitemShut {NoStop}%
\bibitem [{\citenamefont {Christodoulou}\ \emph
  {et~al.}(2023{\natexlab{b}})\citenamefont {Christodoulou}, \citenamefont
  {Di~Biagio}, \citenamefont {Aspelmeyer}, \citenamefont {Brukner},
  \citenamefont {Rovelli},\ and\ \citenamefont
  {Howl}}]{christodoulou2023locally}%
  \BibitemOpen
  \bibfield  {author} {\bibinfo {author} {\bibfnamefont {M.}~\bibnamefont
  {Christodoulou}}, \bibinfo {author} {\bibfnamefont {A.}~\bibnamefont
  {Di~Biagio}}, \bibinfo {author} {\bibfnamefont {M.}~\bibnamefont
  {Aspelmeyer}}, \bibinfo {author} {\bibfnamefont {{\v{C}}.}~\bibnamefont
  {Brukner}}, \bibinfo {author} {\bibfnamefont {C.}~\bibnamefont {Rovelli}},\
  and\ \bibinfo {author} {\bibfnamefont {R.}~\bibnamefont {Howl}},\ }\bibfield
  {title} {\bibinfo {title} {Locally mediated entanglement in linearized
  quantum gravity},\ }\href@noop {} {\bibfield  {journal} {\bibinfo  {journal}
  {Physical Review Letters}\ }\textbf {\bibinfo {volume} {130}},\ \bibinfo
  {pages} {100202} (\bibinfo {year} {2023}{\natexlab{b}})}\BibitemShut
  {NoStop}%
\bibitem [{\citenamefont {Feng}\ \emph {et~al.}(2023)\citenamefont {Feng},
  \citenamefont {Marletto},\ and\ \citenamefont
  {Vedral}}]{feng2023conservation}%
  \BibitemOpen
  \bibfield  {author} {\bibinfo {author} {\bibfnamefont {T.}~\bibnamefont
  {Feng}}, \bibinfo {author} {\bibfnamefont {C.}~\bibnamefont {Marletto}},\
  and\ \bibinfo {author} {\bibfnamefont {V.}~\bibnamefont {Vedral}},\
  }\bibfield  {title} {\bibinfo {title} {Conservation laws reveal the
  quantumness of gravity}} (\bibinfo {year} {2023}),\ \bibinfo {note} {arXiv
  2311.08971}\BibitemShut {NoStop}%
\bibitem [{\citenamefont {Bose}\ \emph {et~al.}(2022)\citenamefont {Bose},
  \citenamefont {Mazumdar}, \citenamefont {Schut},\ and\ \citenamefont
  {Toro{\v{s}}}}]{bose2022mechanism}%
  \BibitemOpen
  \bibfield  {author} {\bibinfo {author} {\bibfnamefont {S.}~\bibnamefont
  {Bose}}, \bibinfo {author} {\bibfnamefont {A.}~\bibnamefont {Mazumdar}},
  \bibinfo {author} {\bibfnamefont {M.}~\bibnamefont {Schut}},\ and\ \bibinfo
  {author} {\bibfnamefont {M.}~\bibnamefont {Toro{\v{s}}}},\ }\bibfield
  {title} {\bibinfo {title} {Mechanism for the quantum natured gravitons to
  entangle masses},\ }\href@noop {} {\bibfield  {journal} {\bibinfo  {journal}
  {Physical Review D}\ }\textbf {\bibinfo {volume} {105}},\ \bibinfo {pages}
  {106028} (\bibinfo {year} {2022})}\BibitemShut {NoStop}%
\bibitem [{\citenamefont {Kafri}\ \emph {et~al.}(2014)\citenamefont {Kafri},
  \citenamefont {Taylor},\ and\ \citenamefont {Milburn}}]{kafri2014classical}%
  \BibitemOpen
  \bibfield  {author} {\bibinfo {author} {\bibfnamefont {D.}~\bibnamefont
  {Kafri}}, \bibinfo {author} {\bibfnamefont {J.}~\bibnamefont {Taylor}},\ and\
  \bibinfo {author} {\bibfnamefont {G.}~\bibnamefont {Milburn}},\ }\bibfield
  {title} {\bibinfo {title} {A classical channel model for gravitational
  decoherence},\ }\href@noop {} {\bibfield  {journal} {\bibinfo  {journal} {New
  Journal of Physics}\ }\textbf {\bibinfo {volume} {16}},\ \bibinfo {pages}
  {065020} (\bibinfo {year} {2014})}\BibitemShut {NoStop}%
\bibitem [{\citenamefont {Khosla}\ and\ \citenamefont
  {Nimmrichter}(2018)}]{Khosla2018}%
  \BibitemOpen
  \bibfield  {author} {\bibinfo {author} {\bibfnamefont {K.~E.}\ \bibnamefont
  {Khosla}}\ and\ \bibinfo {author} {\bibfnamefont {S.}~\bibnamefont
  {Nimmrichter}},\ }\bibfield  {title} {\bibinfo {title} {Classical {Channel}
  {Gravity} in the {Newtonian} {Limit}}} (\bibinfo {year} {2018}),\ \bibinfo
  {note} {arXiv 1812.03118}\BibitemShut {NoStop}%
\bibitem [{\citenamefont
  {Oppenheim}(2023{\natexlab{b}})}]{oppenheim2023postquantum}%
  \BibitemOpen
  \bibfield  {author} {\bibinfo {author} {\bibfnamefont {J.}~\bibnamefont
  {Oppenheim}},\ }\bibfield  {title} {\bibinfo {title} {A postquantum theory of
  classical gravity?},\ }\href@noop {} {\bibfield  {journal} {\bibinfo
  {journal} {Physical Review X}\ }\textbf {\bibinfo {volume} {13}},\ \bibinfo
  {pages} {041040} (\bibinfo {year} {2023}{\natexlab{b}})}\BibitemShut
  {NoStop}%
\bibitem [{\citenamefont {Carney}\ and\ \citenamefont
  {Taylor}(2023)}]{Carney2023}%
  \BibitemOpen
  \bibfield  {author} {\bibinfo {author} {\bibfnamefont {D.}~\bibnamefont
  {Carney}}\ and\ \bibinfo {author} {\bibfnamefont {J.~M.}\ \bibnamefont
  {Taylor}},\ }\bibfield  {title} {\bibinfo {title} {Strongly incoherent
  gravity}} (\bibinfo {year} {2023}),\ \bibinfo {note} {arXiv
  2301.08378}\BibitemShut {NoStop}%
\bibitem [{\citenamefont {Di{\'o}si}(1989)}]{diosi1989models}%
  \BibitemOpen
  \bibfield  {author} {\bibinfo {author} {\bibfnamefont {L.}~\bibnamefont
  {Di{\'o}si}},\ }\bibfield  {title} {\bibinfo {title} {Models for universal
  reduction of macroscopic quantum fluctuations},\ }\href@noop {} {\bibfield
  {journal} {\bibinfo  {journal} {Physical Review A}\ }\textbf {\bibinfo
  {volume} {40}},\ \bibinfo {pages} {1165} (\bibinfo {year}
  {1989})}\BibitemShut {NoStop}%
\bibitem [{\citenamefont {Tilloy}\ and\ \citenamefont
  {Di{\'o}si}(2016)}]{tilloy2016sourcing}%
  \BibitemOpen
  \bibfield  {author} {\bibinfo {author} {\bibfnamefont {A.}~\bibnamefont
  {Tilloy}}\ and\ \bibinfo {author} {\bibfnamefont {L.}~\bibnamefont
  {Di{\'o}si}},\ }\bibfield  {title} {\bibinfo {title} {Sourcing semiclassical
  gravity from spontaneously localized quantum matter},\ }\href@noop {}
  {\bibfield  {journal} {\bibinfo  {journal} {Physical Review D}\ }\textbf
  {\bibinfo {volume} {93}},\ \bibinfo {pages} {024026} (\bibinfo {year}
  {2016})}\BibitemShut {NoStop}%
\bibitem [{\citenamefont {Altamirano}\ \emph {et~al.}(2018)\citenamefont
  {Altamirano}, \citenamefont {Corona-Ugalde}, \citenamefont {Mann},\ and\
  \citenamefont {Zych}}]{altamirano2018gravity}%
  \BibitemOpen
  \bibfield  {author} {\bibinfo {author} {\bibfnamefont {N.}~\bibnamefont
  {Altamirano}}, \bibinfo {author} {\bibfnamefont {P.}~\bibnamefont
  {Corona-Ugalde}}, \bibinfo {author} {\bibfnamefont {R.~B.}\ \bibnamefont
  {Mann}},\ and\ \bibinfo {author} {\bibfnamefont {M.}~\bibnamefont {Zych}},\
  }\bibfield  {title} {\bibinfo {title} {Gravity is not a pairwise local
  classical channel},\ }\href@noop {} {\bibfield  {journal} {\bibinfo
  {journal} {Classical and Quantum Gravity}\ }\textbf {\bibinfo {volume}
  {35}},\ \bibinfo {pages} {145005} (\bibinfo {year} {2018})}\BibitemShut
  {NoStop}%
\bibitem [{\citenamefont {Hall}\ and\ \citenamefont
  {Reginatto}(2018)}]{hall2018two}%
  \BibitemOpen
  \bibfield  {author} {\bibinfo {author} {\bibfnamefont {M.~J.}\ \bibnamefont
  {Hall}}\ and\ \bibinfo {author} {\bibfnamefont {M.}~\bibnamefont
  {Reginatto}},\ }\bibfield  {title} {\bibinfo {title} {On two recent proposals
  for witnessing nonclassical gravity},\ }\href@noop {} {\bibfield  {journal}
  {\bibinfo  {journal} {Journal of Physics A: Mathematical and Theoretical}\
  }\textbf {\bibinfo {volume} {51}},\ \bibinfo {pages} {085303} (\bibinfo
  {year} {2018})}\BibitemShut {NoStop}%
\bibitem [{\citenamefont {Anastopoulos}\ \emph {et~al.}(2021)\citenamefont
  {Anastopoulos}, \citenamefont {Lagouvardos},\ and\ \citenamefont
  {Savvidou}}]{anastopoulos2021gravitational}%
  \BibitemOpen
  \bibfield  {author} {\bibinfo {author} {\bibfnamefont {C.}~\bibnamefont
  {Anastopoulos}}, \bibinfo {author} {\bibfnamefont {M.}~\bibnamefont
  {Lagouvardos}},\ and\ \bibinfo {author} {\bibfnamefont {K.}~\bibnamefont
  {Savvidou}},\ }\bibfield  {title} {\bibinfo {title} {Gravitational effects in
  macroscopic quantum systems: a first-principles analysis},\ }\href@noop {}
  {\bibfield  {journal} {\bibinfo  {journal} {Classical and Quantum Gravity}\
  }\textbf {\bibinfo {volume} {38}},\ \bibinfo {pages} {155012} (\bibinfo
  {year} {2021})}\BibitemShut {NoStop}%
\bibitem [{\citenamefont {Fragkos}\ \emph {et~al.}(2022)\citenamefont
  {Fragkos}, \citenamefont {Kopp},\ and\ \citenamefont
  {Pikovski}}]{fragkos2022inference}%
  \BibitemOpen
  \bibfield  {author} {\bibinfo {author} {\bibfnamefont {V.}~\bibnamefont
  {Fragkos}}, \bibinfo {author} {\bibfnamefont {M.}~\bibnamefont {Kopp}},\ and\
  \bibinfo {author} {\bibfnamefont {I.}~\bibnamefont {Pikovski}},\ }\bibfield
  {title} {\bibinfo {title} {On inference of quantization from gravitationally
  induced entanglement},\ }\href@noop {} {\bibfield  {journal} {\bibinfo
  {journal} {AVS Quantum Science}\ }\textbf {\bibinfo {volume} {4}} (\bibinfo
  {year} {2022})}\BibitemShut {NoStop}%
\bibitem [{\citenamefont {D{\"o}ner}\ and\ \citenamefont
  {Gro{\ss}ardt}(2022)}]{doner2022gravitational}%
  \BibitemOpen
  \bibfield  {author} {\bibinfo {author} {\bibfnamefont {M.~K.}\ \bibnamefont
  {D{\"o}ner}}\ and\ \bibinfo {author} {\bibfnamefont {A.}~\bibnamefont
  {Gro{\ss}ardt}},\ }\bibfield  {title} {\bibinfo {title} {Is gravitational
  entanglement evidence for the quantization of spacetime?},\ }\href@noop {}
  {\bibfield  {journal} {\bibinfo  {journal} {Foundations of Physics}\ }\textbf
  {\bibinfo {volume} {52}},\ \bibinfo {pages} {101} (\bibinfo {year}
  {2022})}\BibitemShut {NoStop}%
\bibitem [{\citenamefont {Spaventa}\ \emph {et~al.}(2023)\citenamefont
  {Spaventa}, \citenamefont {Lami},\ and\ \citenamefont
  {Plenio}}]{spaventa2023tests}%
  \BibitemOpen
  \bibfield  {author} {\bibinfo {author} {\bibfnamefont {G.}~\bibnamefont
  {Spaventa}}, \bibinfo {author} {\bibfnamefont {L.}~\bibnamefont {Lami}},\
  and\ \bibinfo {author} {\bibfnamefont {M.~B.}\ \bibnamefont {Plenio}},\
  }\bibfield  {title} {\bibinfo {title} {On tests of the quantum nature of
  gravitational interactions in presence of non-linear corrections to quantum
  mechanics}} (\bibinfo {year} {2023}),\ \bibinfo {note} {arXiv
  2302.00365}\BibitemShut {NoStop}%
\bibitem [{\citenamefont {Foo}\ \emph {et~al.}(2023)\citenamefont {Foo},
  \citenamefont {Mann},\ and\ \citenamefont {Zych}}]{foo2023relativity}%
  \BibitemOpen
  \bibfield  {author} {\bibinfo {author} {\bibfnamefont {J.}~\bibnamefont
  {Foo}}, \bibinfo {author} {\bibfnamefont {R.~B.}\ \bibnamefont {Mann}},\ and\
  \bibinfo {author} {\bibfnamefont {M.}~\bibnamefont {Zych}},\ }\bibfield
  {title} {\bibinfo {title} {Relativity and decoherence of spacetime
  superpositions}} (\bibinfo {year} {2023}),\ \bibinfo {note} {arXiv
  2302.03259}\BibitemShut {NoStop}%
\bibitem [{\citenamefont {Franson}(2011)}]{franson2011entanglement}%
  \BibitemOpen
  \bibfield  {author} {\bibinfo {author} {\bibfnamefont {J.}~\bibnamefont
  {Franson}},\ }\bibfield  {title} {\bibinfo {title} {Entanglement from
  longitudinal and scalar photons},\ }\href
  {https://doi.org/10.1103/PhysRevA.84.033809} {\bibfield  {journal} {\bibinfo
  {journal} {Physical Review A}\ }\textbf {\bibinfo {volume} {84}},\ \bibinfo
  {pages} {033809} (\bibinfo {year} {2011})}\BibitemShut {NoStop}%
\bibitem [{\citenamefont {Feynman}(1987)}]{Feynman1987-FEYNP}%
  \BibitemOpen
  \bibfield  {author} {\bibinfo {author} {\bibfnamefont {R.~P.}\ \bibnamefont
  {Feynman}},\ }\bibfield  {title} {\bibinfo {title} {Negative probability},\
  }in\ \href@noop {} {\emph {\bibinfo {booktitle} {Quantum Implications: Essays
  in Honour of David Bohm}}},\ \bibinfo {editor} {edited by\ \bibinfo {editor}
  {\bibfnamefont {B.~J.}\ \bibnamefont {Hiley}}\ and\ \bibinfo {editor}
  {\bibfnamefont {D.}~\bibnamefont {Peat}}}\ (\bibinfo  {publisher} {Methuen},\
  \bibinfo {year} {1987})\ pp.\ \bibinfo {pages} {235--248}\BibitemShut
  {NoStop}%
\bibitem [{\citenamefont {Brunner}\ \emph {et~al.}(2014)\citenamefont
  {Brunner}, \citenamefont {Cavalcanti}, \citenamefont {Pironio}, \citenamefont
  {Scarani},\ and\ \citenamefont {Wehner}}]{brunner2014bell}%
  \BibitemOpen
  \bibfield  {author} {\bibinfo {author} {\bibfnamefont {N.}~\bibnamefont
  {Brunner}}, \bibinfo {author} {\bibfnamefont {D.}~\bibnamefont {Cavalcanti}},
  \bibinfo {author} {\bibfnamefont {S.}~\bibnamefont {Pironio}}, \bibinfo
  {author} {\bibfnamefont {V.}~\bibnamefont {Scarani}},\ and\ \bibinfo {author}
  {\bibfnamefont {S.}~\bibnamefont {Wehner}},\ }\bibfield  {title} {\bibinfo
  {title} {Bell nonlocality},\ }\href@noop {} {\bibfield  {journal} {\bibinfo
  {journal} {Reviews of modern physics}\ }\textbf {\bibinfo {volume} {86}},\
  \bibinfo {pages} {419} (\bibinfo {year} {2014})}\BibitemShut {NoStop}%
\bibitem [{\citenamefont {Onggadinata}\ \emph {et~al.}(2023)\citenamefont
  {Onggadinata}, \citenamefont {Kurzynski},\ and\ \citenamefont
  {Kaszlikowski}}]{onggadinata2023simulations}%
  \BibitemOpen
  \bibfield  {author} {\bibinfo {author} {\bibfnamefont {K.}~\bibnamefont
  {Onggadinata}}, \bibinfo {author} {\bibfnamefont {P.}~\bibnamefont
  {Kurzynski}},\ and\ \bibinfo {author} {\bibfnamefont {D.}~\bibnamefont
  {Kaszlikowski}},\ }\bibfield  {title} {\bibinfo {title} {Simulations of
  quantum nonlocality with local negative bits},\ }\href@noop {} {\bibfield
  {journal} {\bibinfo  {journal} {Physical Review A}\ }\textbf {\bibinfo
  {volume} {108}},\ \bibinfo {pages} {032204} (\bibinfo {year}
  {2023})}\BibitemShut {NoStop}%
\bibitem [{\citenamefont {Pl{\'a}vala}(2023)}]{plavala2023general}%
  \BibitemOpen
  \bibfield  {author} {\bibinfo {author} {\bibfnamefont {M.}~\bibnamefont
  {Pl{\'a}vala}},\ }\bibfield  {title} {\bibinfo {title} {General probabilistic
  theories: An introduction},\ }\href@noop {} {\bibfield  {journal} {\bibinfo
  {journal} {Physics Reports}\ }\textbf {\bibinfo {volume} {1033}},\ \bibinfo
  {pages} {1} (\bibinfo {year} {2023})}\BibitemShut {NoStop}%
\bibitem [{\citenamefont {Kochen}\ and\ \citenamefont
  {Specker}(1990)}]{kochen1990problem}%
  \BibitemOpen
  \bibfield  {author} {\bibinfo {author} {\bibfnamefont {S.}~\bibnamefont
  {Kochen}}\ and\ \bibinfo {author} {\bibfnamefont {E.~P.}\ \bibnamefont
  {Specker}},\ }\bibfield  {title} {\bibinfo {title} {The problem of hidden
  variables in quantum mechanics},\ }\href@noop {} {\bibfield  {journal}
  {\bibinfo  {journal} {Ernst Specker Selecta}\ ,\ \bibinfo {pages} {235}}
  (\bibinfo {year} {1990})}\BibitemShut {NoStop}%
\bibitem [{\citenamefont {Spekkens}(2005)}]{spekkens2005contextuality}%
  \BibitemOpen
  \bibfield  {author} {\bibinfo {author} {\bibfnamefont {R.~W.}\ \bibnamefont
  {Spekkens}},\ }\bibfield  {title} {\bibinfo {title} {Contextuality for
  preparations, transformations, and unsharp measurements},\ }\href@noop {}
  {\bibfield  {journal} {\bibinfo  {journal} {Physical Review A}\ }\textbf
  {\bibinfo {volume} {71}},\ \bibinfo {pages} {052108} (\bibinfo {year}
  {2005})}\BibitemShut {NoStop}%
\bibitem [{\citenamefont {Abramsky}\ and\ \citenamefont
  {Brandenburger}(2011)}]{abramsky2011sheaf}%
  \BibitemOpen
  \bibfield  {author} {\bibinfo {author} {\bibfnamefont {S.}~\bibnamefont
  {Abramsky}}\ and\ \bibinfo {author} {\bibfnamefont {A.}~\bibnamefont
  {Brandenburger}},\ }\bibfield  {title} {\bibinfo {title} {The sheaf-theoretic
  structure of non-locality and contextuality},\ }\href@noop {} {\bibfield
  {journal} {\bibinfo  {journal} {New Journal of Physics}\ }\textbf {\bibinfo
  {volume} {13}},\ \bibinfo {pages} {113036} (\bibinfo {year}
  {2011})}\BibitemShut {NoStop}%
\bibitem [{\citenamefont {Raussendorf}(2013)}]{raussendorf2013contextuality}%
  \BibitemOpen
  \bibfield  {author} {\bibinfo {author} {\bibfnamefont {R.}~\bibnamefont
  {Raussendorf}},\ }\bibfield  {title} {\bibinfo {title} {Contextuality in
  measurement-based quantum computation},\ }\href@noop {} {\bibfield  {journal}
  {\bibinfo  {journal} {Physical Review A}\ }\textbf {\bibinfo {volume} {88}},\
  \bibinfo {pages} {022322} (\bibinfo {year} {2013})}\BibitemShut {NoStop}%
\bibitem [{\citenamefont {Dzhafarov}\ \emph {et~al.}(2017)\citenamefont
  {Dzhafarov}, \citenamefont {Cervantes},\ and\ \citenamefont
  {Kujala}}]{dzhafarov2017contextuality}%
  \BibitemOpen
  \bibfield  {author} {\bibinfo {author} {\bibfnamefont {E.~N.}\ \bibnamefont
  {Dzhafarov}}, \bibinfo {author} {\bibfnamefont {V.~H.}\ \bibnamefont
  {Cervantes}},\ and\ \bibinfo {author} {\bibfnamefont {J.~V.}\ \bibnamefont
  {Kujala}},\ }\bibfield  {title} {\bibinfo {title} {Contextuality in canonical
  systems of random variables},\ }\href@noop {} {\bibfield  {journal} {\bibinfo
   {journal} {Philosophical Transactions of the Royal Society A: Mathematical,
  Physical and Engineering Sciences}\ }\textbf {\bibinfo {volume} {375}},\
  \bibinfo {pages} {20160389} (\bibinfo {year} {2017})}\BibitemShut {NoStop}%
\bibitem [{\citenamefont {Mansfield}\ and\ \citenamefont
  {Kashefi}(2018)}]{mansfield2018quantum}%
  \BibitemOpen
  \bibfield  {author} {\bibinfo {author} {\bibfnamefont {S.}~\bibnamefont
  {Mansfield}}\ and\ \bibinfo {author} {\bibfnamefont {E.}~\bibnamefont
  {Kashefi}},\ }\bibfield  {title} {\bibinfo {title} {Quantum advantage from
  sequential-transformation contextuality},\ }\href@noop {} {\bibfield
  {journal} {\bibinfo  {journal} {Physical Review Letters}\ }\textbf {\bibinfo
  {volume} {121}},\ \bibinfo {pages} {230401} (\bibinfo {year}
  {2018})}\BibitemShut {NoStop}%
\bibitem [{\citenamefont {Budroni}\ \emph {et~al.}(2022)\citenamefont
  {Budroni}, \citenamefont {Cabello}, \citenamefont {G{\"u}hne}, \citenamefont
  {Kleinmann},\ and\ \citenamefont {Larsson}}]{budroni2022kochen}%
  \BibitemOpen
  \bibfield  {author} {\bibinfo {author} {\bibfnamefont {C.}~\bibnamefont
  {Budroni}}, \bibinfo {author} {\bibfnamefont {A.}~\bibnamefont {Cabello}},
  \bibinfo {author} {\bibfnamefont {O.}~\bibnamefont {G{\"u}hne}}, \bibinfo
  {author} {\bibfnamefont {M.}~\bibnamefont {Kleinmann}},\ and\ \bibinfo
  {author} {\bibfnamefont {J.-{\AA}.}\ \bibnamefont {Larsson}},\ }\bibfield
  {title} {\bibinfo {title} {Kochen-specker contextuality},\ }\href@noop {}
  {\bibfield  {journal} {\bibinfo  {journal} {Reviews of Modern Physics}\
  }\textbf {\bibinfo {volume} {94}},\ \bibinfo {pages} {045007} (\bibinfo
  {year} {2022})}\BibitemShut {NoStop}%
\bibitem [{\citenamefont {Koopman}(1931)}]{koopman1931hamiltonian}%
  \BibitemOpen
  \bibfield  {author} {\bibinfo {author} {\bibfnamefont {B.~O.}\ \bibnamefont
  {Koopman}},\ }\bibfield  {title} {\bibinfo {title} {Hamiltonian systems and
  transformation in hilbert space},\ }\href@noop {} {\bibfield  {journal}
  {\bibinfo  {journal} {Proceedings of the National Academy of Sciences}\
  }\textbf {\bibinfo {volume} {17}},\ \bibinfo {pages} {315} (\bibinfo {year}
  {1931})}\BibitemShut {NoStop}%
\bibitem [{\citenamefont
  {Neumann}(1932{\natexlab{a}})}]{neumann1932operatorenmethode}%
  \BibitemOpen
  \bibfield  {author} {\bibinfo {author} {\bibfnamefont {J.~v.}\ \bibnamefont
  {Neumann}},\ }\bibfield  {title} {\bibinfo {title} {Zur operatorenmethode in
  der klassischen mechanik},\ }\href@noop {} {\bibfield  {journal} {\bibinfo
  {journal} {Annals of Mathematics}\ ,\ \bibinfo {pages} {587}} (\bibinfo
  {year} {1932}{\natexlab{a}})}\BibitemShut {NoStop}%
\bibitem [{\citenamefont {Neumann}(1932{\natexlab{b}})}]{neumann1932zusatze}%
  \BibitemOpen
  \bibfield  {author} {\bibinfo {author} {\bibfnamefont {J.}~\bibnamefont
  {Neumann}},\ }\bibfield  {title} {\bibinfo {title} {Zusatze zur arbeit, zur
  operatorenmethode in der klassischen mechanik},\ }\href@noop {} {\bibfield
  {journal} {\bibinfo  {journal} {Annals of Mathematics}\ ,\ \bibinfo {pages}
  {789}} (\bibinfo {year} {1932}{\natexlab{b}})}\BibitemShut {NoStop}%
\bibitem [{\citenamefont {Piasecki}(2021)}]{piasecki2021introduction}%
  \BibitemOpen
  \bibfield  {author} {\bibinfo {author} {\bibfnamefont {D.}~\bibnamefont
  {Piasecki}},\ }\bibfield  {title} {\bibinfo {title} {{Introduction to
  Koopman-von Neumann Mechanics}},\ }\Eprint {https://arxiv.org/abs/2112.05619}
  {arXiv:2112.05619}  (\bibinfo {year} {2021})\BibitemShut {NoStop}%
\bibitem [{\citenamefont {Moyal}(1949)}]{moyal1949quantum}%
  \BibitemOpen
  \bibfield  {author} {\bibinfo {author} {\bibfnamefont {J.~E.}\ \bibnamefont
  {Moyal}},\ }\bibfield  {title} {\bibinfo {title} {Quantum mechanics as a
  statistical theory},\ }in\ \href@noop {} {\emph {\bibinfo {booktitle}
  {Mathematical Proceedings of the Cambridge Philosophical Society}}},\
  Vol.~\bibinfo {volume} {45}\ (\bibinfo {organization} {Cambridge University
  Press},\ \bibinfo {year} {1949})\ pp.\ \bibinfo {pages} {99--124}\BibitemShut
  {NoStop}%
\bibitem [{\citenamefont {Groenewold}(1946)}]{groenewold1946principles}%
  \BibitemOpen
  \bibfield  {author} {\bibinfo {author} {\bibfnamefont {H.~J.}\ \bibnamefont
  {Groenewold}},\ }\href@noop {} {\emph {\bibinfo {title} {On the principles of
  elementary quantum mechanics}}}\ (\bibinfo  {publisher} {Springer},\ \bibinfo
  {year} {1946})\BibitemShut {NoStop}%
\bibitem [{\citenamefont {Case}(2008)}]{case2008wigner}%
  \BibitemOpen
  \bibfield  {author} {\bibinfo {author} {\bibfnamefont {W.~B.}\ \bibnamefont
  {Case}},\ }\bibfield  {title} {\bibinfo {title} {Wigner functions and weyl
  transforms for pedestrians},\ }\href@noop {} {\bibfield  {journal} {\bibinfo
  {journal} {American Journal of Physics}\ }\textbf {\bibinfo {volume} {76}},\
  \bibinfo {pages} {937} (\bibinfo {year} {2008})}\BibitemShut {NoStop}%
\bibitem [{\citenamefont {Schleich}(2011)}]{schleich2011quantum}%
  \BibitemOpen
  \bibfield  {author} {\bibinfo {author} {\bibfnamefont {W.~P.}\ \bibnamefont
  {Schleich}},\ }\href@noop {} {\emph {\bibinfo {title} {Quantum optics in
  phase space}}}\ (\bibinfo  {publisher} {John Wiley \& Sons},\ \bibinfo {year}
  {2011})\BibitemShut {NoStop}%
\bibitem [{\citenamefont {Pl{\'a}vala}\ and\ \citenamefont
  {Kleinmann}(2022)}]{plavala2022operational}%
  \BibitemOpen
  \bibfield  {author} {\bibinfo {author} {\bibfnamefont {M.}~\bibnamefont
  {Pl{\'a}vala}}\ and\ \bibinfo {author} {\bibfnamefont {M.}~\bibnamefont
  {Kleinmann}},\ }\bibfield  {title} {\bibinfo {title} {Operational theories in
  phase space: Toy model for the harmonic oscillator},\ }\href@noop {}
  {\bibfield  {journal} {\bibinfo  {journal} {Physical Review Letters}\
  }\textbf {\bibinfo {volume} {128}},\ \bibinfo {pages} {040405} (\bibinfo
  {year} {2022})}\BibitemShut {NoStop}%
\bibitem [{\citenamefont {Almeida}(2009)}]{almeida2009entanglement}%
  \BibitemOpen
  \bibfield  {author} {\bibinfo {author} {\bibfnamefont {A.~O.~d.}\
  \bibnamefont {Almeida}},\ }\bibfield  {title} {\bibinfo {title} {Entanglement
  in phase space},\ }in\ \href@noop {} {\emph {\bibinfo {booktitle}
  {Entanglement and Decoherence: Foundations and Modern Trends}}}\ (\bibinfo
  {publisher} {Springer},\ \bibinfo {year} {2009})\ pp.\ \bibinfo {pages}
  {157--219}\BibitemShut {NoStop}%
\bibitem [{\citenamefont {Pl{\'a}vala}\ and\ \citenamefont
  {Kleinmann}(2023)}]{plavala2023generalized}%
  \BibitemOpen
  \bibfield  {author} {\bibinfo {author} {\bibfnamefont {M.}~\bibnamefont
  {Pl{\'a}vala}}\ and\ \bibinfo {author} {\bibfnamefont {M.}~\bibnamefont
  {Kleinmann}},\ }\bibfield  {title} {\bibinfo {title} {Generalized dynamical
  theories in phase space and the hydrogen atom},\ }\href@noop {} {\bibfield
  {journal} {\bibinfo  {journal} {Physical Review A}\ }\textbf {\bibinfo
  {volume} {108}},\ \bibinfo {pages} {052212} (\bibinfo {year}
  {2023})}\BibitemShut {NoStop}%
\bibitem [{\citenamefont {Renou}\ \emph {et~al.}(2021)\citenamefont {Renou},
  \citenamefont {Trillo}, \citenamefont {Weilenmann}, \citenamefont {Le},
  \citenamefont {Tavakoli}, \citenamefont {Gisin}, \citenamefont {Ac{\'\i}n},\
  and\ \citenamefont {Navascu{\'e}s}}]{renou2021quantum}%
  \BibitemOpen
  \bibfield  {author} {\bibinfo {author} {\bibfnamefont {M.-O.}\ \bibnamefont
  {Renou}}, \bibinfo {author} {\bibfnamefont {D.}~\bibnamefont {Trillo}},
  \bibinfo {author} {\bibfnamefont {M.}~\bibnamefont {Weilenmann}}, \bibinfo
  {author} {\bibfnamefont {T.~P.}\ \bibnamefont {Le}}, \bibinfo {author}
  {\bibfnamefont {A.}~\bibnamefont {Tavakoli}}, \bibinfo {author}
  {\bibfnamefont {N.}~\bibnamefont {Gisin}}, \bibinfo {author} {\bibfnamefont
  {A.}~\bibnamefont {Ac{\'\i}n}},\ and\ \bibinfo {author} {\bibfnamefont
  {M.}~\bibnamefont {Navascu{\'e}s}},\ }\bibfield  {title} {\bibinfo {title}
  {Quantum theory based on real numbers can be experimentally falsified},\
  }\href@noop {} {\bibfield  {journal} {\bibinfo  {journal} {Nature}\ }\textbf
  {\bibinfo {volume} {600}},\ \bibinfo {pages} {625} (\bibinfo {year}
  {2021})}\BibitemShut {NoStop}%
\bibitem [{\citenamefont {Catani}\ \emph {et~al.}(2022)\citenamefont {Catani},
  \citenamefont {Leifer}, \citenamefont {Scala}, \citenamefont {Schmid},\ and\
  \citenamefont {Spekkens}}]{catani2022nonclassical}%
  \BibitemOpen
  \bibfield  {author} {\bibinfo {author} {\bibfnamefont {L.}~\bibnamefont
  {Catani}}, \bibinfo {author} {\bibfnamefont {M.}~\bibnamefont {Leifer}},
  \bibinfo {author} {\bibfnamefont {G.}~\bibnamefont {Scala}}, \bibinfo
  {author} {\bibfnamefont {D.}~\bibnamefont {Schmid}},\ and\ \bibinfo {author}
  {\bibfnamefont {R.~W.}\ \bibnamefont {Spekkens}},\ }\bibfield  {title}
  {\bibinfo {title} {What is nonclassical about uncertainty relations?},\
  }\href@noop {} {\bibfield  {journal} {\bibinfo  {journal} {Physical Review
  Letters}\ }\textbf {\bibinfo {volume} {129}},\ \bibinfo {pages} {240401}
  (\bibinfo {year} {2022})}\BibitemShut {NoStop}%
\bibitem [{\citenamefont {Blasiak}(2015)}]{blasiak2015local}%
  \BibitemOpen
  \bibfield  {author} {\bibinfo {author} {\bibfnamefont {P.}~\bibnamefont
  {Blasiak}},\ }\bibfield  {title} {\bibinfo {title} {Local model of a qubit in
  the interferometric setup},\ }\href@noop {} {\bibfield  {journal} {\bibinfo
  {journal} {New Journal of Physics}\ }\textbf {\bibinfo {volume} {17}},\
  \bibinfo {pages} {113043} (\bibinfo {year} {2015})}\BibitemShut {NoStop}%
\bibitem [{\citenamefont {Catani}\ \emph
  {et~al.}(2023{\natexlab{a}})\citenamefont {Catani}, \citenamefont {Leifer},
  \citenamefont {Scala}, \citenamefont {Schmid},\ and\ \citenamefont
  {Spekkens}}]{catani2023aspects}%
  \BibitemOpen
  \bibfield  {author} {\bibinfo {author} {\bibfnamefont {L.}~\bibnamefont
  {Catani}}, \bibinfo {author} {\bibfnamefont {M.}~\bibnamefont {Leifer}},
  \bibinfo {author} {\bibfnamefont {G.}~\bibnamefont {Scala}}, \bibinfo
  {author} {\bibfnamefont {D.}~\bibnamefont {Schmid}},\ and\ \bibinfo {author}
  {\bibfnamefont {R.~W.}\ \bibnamefont {Spekkens}},\ }\bibfield  {title}
  {\bibinfo {title} {Aspects of the phenomenology of interference that are
  genuinely nonclassical},\ }\href@noop {} {\bibfield  {journal} {\bibinfo
  {journal} {Physical Review A}\ }\textbf {\bibinfo {volume} {108}},\ \bibinfo
  {pages} {022207} (\bibinfo {year} {2023}{\natexlab{a}})}\BibitemShut
  {NoStop}%
\bibitem [{\citenamefont {Catani}\ \emph
  {et~al.}(2023{\natexlab{b}})\citenamefont {Catani}, \citenamefont {Leifer},
  \citenamefont {Schmid},\ and\ \citenamefont
  {Spekkens}}]{catani2023interference}%
  \BibitemOpen
  \bibfield  {author} {\bibinfo {author} {\bibfnamefont {L.}~\bibnamefont
  {Catani}}, \bibinfo {author} {\bibfnamefont {M.}~\bibnamefont {Leifer}},
  \bibinfo {author} {\bibfnamefont {D.}~\bibnamefont {Schmid}},\ and\ \bibinfo
  {author} {\bibfnamefont {R.~W.}\ \bibnamefont {Spekkens}},\ }\bibfield
  {title} {\bibinfo {title} {Why interference phenomena do not capture the
  essence of quantum theory},\ }\href@noop {} {\bibfield  {journal} {\bibinfo
  {journal} {Quantum}\ }\textbf {\bibinfo {volume} {7}},\ \bibinfo {pages}
  {1119} (\bibinfo {year} {2023}{\natexlab{b}})}\BibitemShut {NoStop}%
\end{thebibliography}
\end{document}